\documentclass[twocolumn]{aastex631}
\usepackage{xspace,amsmath,amssymb,multirow}

\usepackage{afterpage}

\newcommand{\per}{\ensuremath{^{-1}}\xspace}
\newcommand{\Ha}{H\ensuremath{\alpha}\xspace}

\shorttitle{Broad-Line AGN at $3.5<z<6$}
\shortauthors{Taylor et al.}

\begin{document}

\title{Broad-Line AGN at $\mathbf{3.5<z<6}$:\\
The Black Hole Mass Function and a Connection with Little Red Dots}

\correspondingauthor{Anthony~J.~Taylor}
\email{anthony.taylor@austin.utexas.edu}



\author[0000-0003-1282-7454]{Anthony J. Taylor}
\affiliation{Department of Astronomy, The University of Texas at Austin, Austin, TX, USA}

\author[0000-0001-8519-1130]{Steven L. Finkelstein}
\affiliation{Department of Astronomy, The University of Texas at Austin, Austin, TX, USA}

\author[0000-0002-8360-3880]{Dale D. Kocevski}
\affiliation{Department of Physics and Astronomy, Colby College, Waterville, ME 04901, USA}

\author[0000-0002-6038-5016]{Junehyoung Jeon}
\affiliation{Department of Astronomy, The University of Texas at Austin, Austin, TX, USA}

\author[0000-0003-0212-2979]{Volker Bromm}
\affiliation{Department of Astronomy, The University of Texas at Austin, Austin, TX, USA}


\author[0000-0001-5758-1000]{Ricardo O. Amor\'{i}n}
\affiliation{Instituto de Astrof\'{i}sica de Andaluc\'{i}a (CSIC), Apartado 3004, 18080 Granada, Spain}

\author[0000-0002-7959-8783]{Pablo Arrabal Haro}
\affiliation{NSF's National Optical-Infrared Astronomy Research Laboratory, 950 N. Cherry Ave., Tucson, AZ 85719, USA}

\author[0000-0001-8534-7502]{Bren E. Backhaus}
\affiliation{Department of Physics, 196 Auditorium Road, Unit 3046, University of Connecticut, Storrs, CT 06269, USA}

\author[0000-0002-9921-9218]{Micaela B. Bagley}
\affiliation{Department of Astronomy, The University of Texas at Austin, Austin, TX, USA}

\author[0000-0002-2931-7824]{Eduardo Ba\~nados}
\affiliation{{Max-Planck-Institut f\"ur Astronomie, K\"onigstuhl 17, D-69117, Heidelberg, Germany}}

\author[0000-0003-0883-2226]{Rachana Bhatawdekar}
\affiliation{European Space Agency (ESA), European Space Astronomy Centre (ESAC), Camino Bajo del Castillo s/n, 28692 Villanueva de la Cañada, Madrid, Spain}

\author[0000-0001-5384-3616]{Madisyn Brooks}
\affiliation{Department of Physics, 196 Auditorium Road, Unit 3046, University of Connecticut, Storrs, CT 06269, USA}

\author[0000-0003-2536-1614]{Antonello Calabr{\`o}} 
\affiliation{INAF - Osservatorio Astronomico di Roma, via di Frascati 33, 00078 Monte Porzio Catone, Italy}

\author[0000-0003-2332-5505]{\'Oscar A. Ch\'avez Ortiz}
\affiliation{Department of Astronomy, The University of Texas at Austin, Austin, TX, USA}

\author[0000-0001-8551-071X]{Yingjie Cheng}
\affiliation{University of Massachusetts Amherst, 710 North Pleasant Street, Amherst, MA 01003-9305, USA}

\author[0000-0001-7151-009X]{Nikko J. Cleri}
\affiliation{Department of Astronomy and Astrophysics, The Pennsylvania State University, University Park, PA 16802, USA}
\affiliation{Institute for Computational and Data Sciences, The Pennsylvania State University, University Park, PA 16802, USA}
\affiliation{Institute for Gravitation and the Cosmos, The Pennsylvania State University, University Park, PA 16802, USA}

\author[0000-0002-6348-1900]{Justin W. Cole}
\affiliation{Department of Physics and Astronomy, Texas A\&M University, College Station, TX, 77843-4242 USA}
\affiliation{George P.\ and Cynthia Woods Mitchell Institute for Fundamental Physics and Astronomy, Texas A\&M University, College Station, TX, 77843-4242 USA}

\author[0000-0001-8047-8351]{Kelcey Davis}
\affiliation{Department of Physics, 196 Auditorium Road, Unit 3046, University of Connecticut, Storrs, CT 06269, USA}

\author[0000-0001-5414-5131]{Mark Dickinson}
\affiliation{NSF's National Optical-Infrared Astronomy Research Laboratory, 950 N. Cherry Ave., Tucson, AZ 85719, USA}

\author[0000-0002-7622-0208]{Callum Donnan}
\affiliation{Institute for Astronomy, University of Edinburgh, Royal Observatory, Edinburgh EH9 3HJ, UK}

\author[0000-0002-1404-5950]{James S. Dunlop}
\affiliation{SUPA, Institute for Astronomy, University of Edinburgh, Royal Observatory, Edinburgh EH9 3HJ, UK}

\author[0000-0001-7782-7071]{Richard S. Ellis}
\affiliation{Department of Physics and Astronomy, University College London, Gower Street, London WC1E 6BT, UK}

\author[0000-0003-0531-5450]{Vital Fern\'{a}ndez}
\affiliation{Instituto de Investigaci\'{o}n Multidisciplinar en Ciencia y Tecnolog\'{i}a, Universidad de La Serena, Raul Bitr\'{a}n 1305, La Serena 2204000, Chile}

\author[0000-0003-3820-2823]{Adriano Fontana}
\affiliation{INAF - Osservatorio Astronomico di Roma, via di Frascati 33, 00078 Monte Porzio Catone, Italy}

\author[0000-0001-7201-5066]{Seiji Fujimoto}
\affiliation{Department of Astronomy, The University of Texas at Austin, Austin, TX, USA}

\author[0000-0002-7831-8751]{Mauro Giavalisco}
\affiliation{University of Massachusetts Amherst, 710 North Pleasant Street, Amherst, MA 01003-9305, USA}

\author[0000-0002-5688-0663]{Andrea Grazian}
\affiliation{INAF--Osservatorio Astronomico di Padova, Vicolo dell'Osservatorio 5, I-35122, Padova, Italy}

\author{Jingsong Guo}
\affiliation{Department of Astronomy, School of Physics, Peking University, Beijing 100871, China}

\author[0000-0001-6145-5090]{Nimish P. Hathi}
\affiliation{Space Telescope Science Institute, 3700 San Martin Drive, Baltimore, MD 21218, USA}

\author[0000-0002-4884-6756]{Benne W. Holwerda}
\affil{Physics \& Astronomy Department, University of Louisville, 40292 KY, Louisville, USA}

\author[0000-0002-3301-3321]{Michaela Hirschmann}
\affiliation{Institute of Physics, Laboratory of Galaxy Evolution, Ecole Polytechnique Fédérale de Lausanne (EPFL), Observatoire de Sauverny, 1290 Versoix, Switzerland}

\author[0000-0001-9840-4959]{Kohei Inayoshi}
\affiliation{Kavli Institute for Astronomy and Astrophysics, Peking University, Beijing 100871, China}

\author[0000-0001-9187-3605]{Jeyhan S. Kartaltepe}
\affiliation{Laboratory for Multiwavelength Astrophysics, School of Physics and Astronomy, Rochester Institute of Technology, 84 Lomb Memorial Drive, Rochester, NY 14623, USA}

\author[0000-0002-7220-397X]{Yana Khusanova}
\affiliation{Max-Planck-Institut f\"{u}r Astronomie, K\"{o}nigstuhl 17, D-69117 Heidelberg, Germany}

\author[0000-0002-6610-2048]{Anton M. Koekemoer}
\affiliation{Space Telescope Science Institute, 3700 San Martin Drive, Baltimore, MD 21218, USA}

\author[0000-0002-5588-9156]{Vasily Kokorev}
\affiliation{Department of Astronomy, The University of Texas at Austin, Austin, TX, USA}

\author[0000-0003-2366-8858]{Rebecca L. Larson}
\affiliation{NSF Graduate Fellow}
\affiliation{Department of Astronomy, The University of Texas at Austin, Austin, TX, USA}

\author[0000-0002-9393-6507]{Gene C. K. Leung}
\affiliation{Department of Astronomy, The University of Texas at Austin, Austin, TX, USA}
\affiliation{MIT Kavli Institute for Astrophysics and Space Research, 77 Massachusetts Ave., Cambridge, MA 02139, USA}

\author[0000-0003-1581-7825]{Ray A. Lucas}
\affiliation{Space Telescope Science Institute, 3700 San Martin Drive, Baltimore, MD 21218, USA}

\author[0000-0003-4368-3326]{Derek J. McLeod}\affiliation{Institute for Astronomy, University of Edinburgh, Royal Observatory, Edinburgh, EH9 3HJ, UK}

\author[0000-0002-8951-4408]{Lorenzo Napolitano}
\affiliation{INAF – Osservatorio Astronomico di Roma, via Frascati 33, 00078, Monteporzio Catone, Italy}
\affiliation{Dipartimento di Fisica, Università di Roma Sapienza, Città Universitaria di Roma - Sapienza, Piazzale Aldo Moro, 2, 00185, Roma, Italy}

\author[0000-0003-2984-6803]{Masafusa Onoue} 
\altaffiliation{Kavli Astrophysics Fellow}
\affil{Kavli Institute for the Physics and Mathematics of the Universe (Kavli IPMU, WPI), The University of Tokyo Institutes for Advanced Study, The University of Tokyo, Kashiwa, Chiba 277-8583, Japan} 
\affil{Center for Data-Driven Discovery, Kavli IPMU (WPI), UTIAS, The University of Tokyo, Kashiwa, Chiba 277-8583, Japan}
\affil{Kavli Institute for Astronomy and Astrophysics, Peking University, Beijing 100871, P.R.China}

\author[0000-0001-9879-7780]{Fabio Pacucci}
\affiliation{Center for Astrophysics $\vert$ Harvard \& Smithsonian, 60 Garden St, Cambridge, MA 02138, USA}
\affiliation{Black Hole Initiative, Harvard University, 20 Garden St, Cambridge, MA 02138, USA}

\author[0000-0001-7503-8482]{Casey Papovich}
\affiliation{Department of Physics and Astronomy, Texas A\&M University, College Station, TX, 77843-4242 USA}
\affiliation{George P.\ and Cynthia Woods Mitchell Institute for Fundamental Physics and Astronomy, Texas A\&M University, College Station, TX, 77843-4242 USA}

\author[0000-0003-4528-5639]{Pablo G. P\'erez-Gonz\'alez}
\affiliation{Centro de Astrobiolog\'{\i}a (CAB), CSIC-INTA, Ctra. de Ajalvir km 4, Torrej\'on de Ardoz, E-28850, Madrid, Spain}

\author[0000-0003-3382-5941]{Nor Pirzkal}
\affiliation{ESA/AURA Space Telescope Science Institute}

\author[0000-0002-6748-6821]{Rachel S. Somerville}
\affiliation{Center for Computational Astrophysics, Flatiron Institute, 162 5th Avenue, New York, NY, 10010, USA}

\author[0000-0002-1410-0470]{Jonathan R. Trump}
\affiliation{Department of Physics, 196 Auditorium Road, Unit 3046, University of Connecticut, Storrs, CT 06269, USA}

\author[0000-0003-3903-6935]{Stephen M.~Wilkins} %
\affiliation{Astronomy Centre, University of Sussex, Falmer, Brighton BN1 9QH, UK}
\affiliation{Institute of Space Sciences and Astronomy, University of Malta, Msida MSD 2080, Malta}

\author[0000-0003-3466-035X]{{L. Y. Aaron} {Yung}}
\affiliation{Space Telescope Science Institute, 3700 San Martin Drive, Baltimore, MD 21218, USA}

\author[0000-0002-4321-3538]{Haowen Zhang}
\affiliation{Steward Observatory, University of Arizona, 933 N Cherry Ave., Tucson, AZ 85721, USA}

\begin{abstract}
We present a sample of 62 H$\alpha$ detected broad-line active galactic nuclei (BLAGN) at redshifts $3.5<z<6.8$ using data from the CEERS and RUBIES surveys. We select these sources directly from \textit{JWST}/NIRSpec G395M/F290LP spectra. We use a multi-step pre-selection and Bayesian fitting to ensure a high-quality sample of sources with broad Balmer lines and narrow forbidden lines. We compute rest-frame ultraviolet and optical spectral slopes for these objects, and determine that 21 BLAGN in our sample are also little red dots (LRDs). These LRD BLAGN, when examined in aggregate, show broader H$\alpha$ line profiles and a higher fraction of broad-to-narrow component H$\alpha$ emission than non-LRD BLAGN. We find that $\sim90\%$ of LRD BLAGN are intrinsically reddened ($\beta_{opt}>0$), independent of contributions from emission lines to the broadband photometry. We construct the black hole (BH) mass function at $3.5<z<6$ after computing robust completeness corrections. This BH mass function shows agreement with recent \textit{JWST}-based BH mass functions, though we extend these earlier results to log (M$_{BH}/M_{\odot}) < 7$. The derived BH mass function is consistent with theoretical models, indicating that the observed abundance of BHs in the early universe may not be discrepant with physically-motivated predictions. The BH mass function shape resembles a largely featureless power-law, suggesting that signatures from BH seeding have been lost by redshift $z\sim$5--6. Finally, we compute the BLAGN UV luminosity function and find agreement with \textit{JWST}-detected BLAGN samples from recent works, finding that BLAGN hosts constitute $\lesssim$10\% of the total observed UV luminosity at all but the brightest luminosities.
\end{abstract}

\section{Introduction} \label{sec:intro}

It is well known that the majority of galaxies host supermassive black holes (SMBHs) that play critical roles in galaxy evolution across cosmic time. These roles are most evident when these SMBHs actively accrete baryonic material and become active galactic nuclei (AGN; e.g. \citealt{heckman14,greene20}). AGN are classically divided into two categories. In type I AGN, the accretion disk of infalling material is unobscured, allowing an observer to directly probe the emission from the rapidly rotating accretion disk. In type II AGN, the accretion disk is obscured by gas and dust, and signatures from the highly kinetic matter near the central SMBH are hidden. As such, type I AGN are distinctly easier to identify due to their clear highly kinetically broadened emission lines, and are thus often referred to as broad-line AGN (BLAGN). While emission line diagnostics such as the BPT diagram \citep{baldwin81,kewley01,kauffmann03} can successfully distinguish both type I and type II AGN from star-forming galaxies in the local universe, these diagnostics are not yet calibrated nor reliable in the high-redshift universe \citep[e.g.,][]{grag22,brinchmann23,pacucci23,scholtz23,chisholm24,mazzolari24,pacucci24}; although some systems also do appear to match local relations \citep[e.g.,][]{maiolino23b,kokorev24b}. Despite these challenges, observations of these objects are critical to advancing our understanding of galaxy evolution at early cosmic times. 

Studying AGN can help constrain the origin of SMBHs.  Typically, black hole seeds are classified as either light ($\lesssim 10^3 \, \rm M_\odot$) or heavy ($\gtrsim 10^4 \, \rm M_\odot$). Light seeds are formed as remnants of Population~III stars \citep[e.g.,][]{Madau2001}, or via alternative dynamical processes in dense stellar environments (e.g., \citealt{PZ_2002, Freitag_2006, Miller_2012, Katz_2015, Lupi_2016, Boekholt_2018,GonzalezKremer2021}). Heavy seeds\footnote{Another possibility is that heavy seeds could have originated in the ultra-early universe, giving rise to primordial BHs \citep[e.g.][]{LiuBromm2022}}, for example via the direct collapse black hole (DCBH) channel, are formed from the monolithic collapse of metal-free gas clouds at $z\gtrsim 10$ \citep[e.g.,][]{Loeb_Rasio_1994, Bromm_Loeb_2003, Lodato_Natarajan_2006}. While light BH seeds could, through steady Eddington-limited accretion, grow from early times to form the massive SMBHs seen locally, this seeding mechanism cannot yield the observed masses of the most massive SMBHs seen at $z \gtrsim$ 6 \citep[e.g.][]{fan06,yang20}.  Mechanisms such as super-Eddington accretion, or the formation of more massive seeds \citep[e.g.][]{Bromm_Loeb_2003} could in principle form the observed population of the most massive SMBHs at high redshift.  However, disentangling these seeding mechanisms requires measuring the distribution function of black hole masses, requiring constraints on lower-mass BHs at high-redshift. In particular, the high-redshift BH mass function is highly instructive as it may constrain the seeds of the first AGN and quasars \citep[e.g.,][]{Ferrara_2014,Smith2019,Inayoshi2020}.

Prior to the launch of \textit{JWST}, spectroscopic observations of BLAGN in the redshift $z>3$ universe were challenging, primarily due to the observed wavelength of the \Ha line---a dominant tracer of BLAGN---entering the thermal infrared at $\gtrsim2.5$~$\mu$m, and thus becoming difficult to observe from the ground due to the poor transmission of these wavelengths through the Earth's atmosphere, as well as the presence of sky-lines and atmospheric and telescope thermal emission. Nonetheless, photometric and spectroscopic studies assembled large catalogs of the galaxies hosting the most massive and strongly emitting BLAGN at $z>5$ from programs such as the Sloan Digital Sky Survey \citep{york00,fan00,shen12,jiang16,wu22}, Pan-STARRS1 \citep{banados16}, and most recently with the Hyper Suprime-Cam instrument on the Subaru 8.2m Telescope \citep[e.g.][]{he24} using alternative detection methods beyond the \Ha line, such as broad C\,\textsc{iv}$\lambda\lambda1548,1550$ or Mg\,\textsc{ii}]$\lambda\lambda2795,2803$ lines, or color selections. While these surveys were sensitive to BLAGN hosting black holes (BHs) with masses $M_{\rm BH}\gtrsim10^8 M_{\odot}$, they were insufficiently sensitive to detect even lighter BHs in the early $z>3$ universe. 

Beyond offering insight into the evolution of SMBHs, AGN also may play a critical role in cosmic reionization \citep[e.g.,][and references therein]{giallongo15,madau15,finkelstein19,pratika20}. Recent post-\textit{JWST} studies have offered new evidence that AGN may indeed be significant drivers of reionization \citep{madau24}, and may make up a larger fraction of the epoch of reionization UV luminosity function than previously expected \citep[e.g.,][]{grazian24}.

The challenge of detecting these AGN was removed with the launch of the \textit{JWST} in late 2021. In the following years, due primarily to \textit{JWST}'s unprecedented near-infrared sensitivity, enormous progress has been made in detecting BLAGN at $z>3$, resulting in BLAGN detections at $3<z<8$ and beyond \citep[e.g.][]{larson23,harikane23,Maiolino_2023,kocevski23,kocevski24,matthee24}. These studies have successfully detected BLAGN down to $M_{\rm BH}\sim 10^7 M_{\odot}$ in large area surveys such as CEERS \citep{finkelstein17}, UNCOVER \citep{bezanson22}, JADES \citep{eisenstein23}, EIGER \citep{kashino23}, and FRESCO \citep{oesch23}.

Interestingly, in an early sample of 10 BLAGN identified with \textit{JWST}, \cite{harikane23} found that 20\% of these objects appear to be heavily obscured. These objects feature a steep red continuum in the rest-frame optical, while also exhibiting relatively blue colors in the rest-frame UV \citep{kocevski23, matthee24, greene24, Killi23}. Compact sources with this “v-shaped”, red plus blue spectral energy distribution (SED) have come to be known as “little red dots” (LRDs) in the literature \citep{matthee24}.  

While the nature of LRDs, and whether they are powered by AGN, is still heavily debated (e.g., \citealt[][Leung et al., in prep]{Barro24, labbe23, Akins24, PerezGonzalez24, Williams24, Li24, Baggen24}), their unique SEDs and compact morphologies allow them to be identified photometrically, either using color criteria \citep{Barro24, labbe23, kokorev24} or cuts on their rest-UV and optical continuum slopes \citep{kocevski24}. Studies have found LRDs selected these ways to be quite ubiquitous, having number densities of $\sim10^{-5}$ Mpc$^{-3}$ mag$^{-1}$, which amounts to a few percent of the galaxy population at redshift $z\sim5-6$ \citep{kokorev24, kocevski24}.  Spectroscopic observations of photometrically selected LRDs reveals that over 70--80\% show broad-line emission when care is taken to exclude brown dwarfs \citep{greene24, kocevski24}.  If most LRDs are indeed powered by heavily obscured AGN, then they may provide a unique window into the early, obscured growth phase of today's SMBHs.

In this work, we aim to leverage \textit{JWST}'s unprecedented ability to directly detect BLAGN to compile a statistically-significant sample of BLAGN at $z >$ 3 by searching in the CEERS and PRIMER UDS fields using spectra from \textit{JWST}/NIRSpec. We use this sample to construct the BLAGN BH mass function, BLAGN UV luminosity function, and analyze our sample's overlap with populations of LRDs.

We organize this work as follows: In \S\ref{sec:observations} we describe the \textit{JWST} and \textit{HST} photometric and spectroscopic datasets used in this study, including their reduction and calibration. In \S\ref{sec:analysis} and subsections therein, we describe our systematic search for BLAGN using the \Ha line in the above-mentioned \textit{JWST}/NIRSpec data. In \S\ref{sec:sample} we describe the detected BLAGN sample. In \S\ref{sec:lrds}, we compare our BLAGN sample with recent studies of ``Little Red Dots''. In \S\ref{sec:BH mass function}, we construct the BH mass function using our BLAGN sample and compare to recent observational results from both \textit{JWST} and ground-based data, as well as empirical and theoretical literature models including some simple toy models of our own construction. In \S\ref{sec:UVLF} we construct the BLAGN ultraviolet (UV) luminosity function from our sample and again compare to recent observations. Finally, we summarize our work in \S\ref{sec:summary}.

We assume $\Omega_m$=$0.3$, $\Omega_\Lambda$=$0.7$, and $H_0$=$70$~km~s$^{-1}$~Mpc$^{-1}$ throughout. All magnitudes are given in the AB magnitude system, where an AB magnitude is defined by $m_{AB}$=$-2.5\log f_\nu - 48.60$. Here, $f_\nu$ is the specific flux of the source in units of erg~cm$^{-2}$~s$^{-1}$~Hz$^{-1}$. 

\section{Observations}\label{sec:observations}

\subsection{Photometry}\label{sec:photometry}

The analysis in this paper makes use of \textit{JWST}/NIRCam photometry in the CEERS and PRIMER/UDS fields.  The CEERS-field imaging is from the CEERS team's v1.0 reduction (Finkelstein et al., in prep; these data will soon be available on the CEERS website), and are supplemented by F090W data on the same field from \textit{JWST} GO\#2234, (PI Banados).  The imaging in the UDS field comes from the PRIMER survey (PI Dunlop), and is an internal reduction from the PRIMER team (internal version 0.6). The PRIMER imaging data were reduced using the PRIMER Enhanced NIRCam Image Processing Library (PENCIL; Magee et al., in preparation, Dunlop et al. in preparation) software. The astrometry of all the reduced images was aligned to GAIA Data Release 3 \citep{gaia23} and stacked to the same pixel scale of 0\farcs03.  Both fields contain the same set of NIRCam filters (F090W, F115W, F150W, F200W, F277W, F356W, F410M and F444W).  We also make use of \textit{HST}/ACS F606W and F814W imaging in both fields \citep{grogin11,koekemoer11}.  The photometry catalogs also include \textit{HST}/WFC3 F105W, F125W, F140W and F160W information, but as this is superseded by the NIRCam imaging, these bands are not used in the analysis.

The photometry in both fields was calculated following the process outlined in \citet{finkelstein24}.  In brief, this methodology focuses on measuring accurate colors and total flux estimates across \textit{HST}/ACS, WFC3 and \textit{JWST}/NIRCam imaging.  Accurate colors are achieved by PSF-matching images (using empirical PSFs) with smaller PSFs than F277W to that band, and deriving correction factors for images with larger PSFs (via PSF-matching F277W to a given larger PSF).  Small Kron apertures are used to measure colors to optimize signal-to-noise for high-redshift galaxies.  Total fluxes are estimated by first deriving an aperture correction in the F277W band as the ratio between the flux in the larger (default) Kron aperture and the custom smaller aperture, with a residual aperture correction (typically $<$10\%) derived via source-injection simulations.

The key difference between the procedure described in \citet{finkelstein24} and that used here is the inclusion of a ``hot$+$cold'' step, where first sources are selected with conservative (cold) detection parameters, designed to not split up large, bright galaxies (qualitatively similar to the process done in \citealt{galametz13}).  Then, a more aggressive (hot) run is performed to identify fainter objects.  Objects from the hot catalog that fall outside the segmentation map from the cold run are added to the final photometry catalog.  

Catalogs were created in both the CEERS and UDS fields in an identical manner. Photometric redshifts were estimated with \texttt{EAZY} \citep{brammer08}, using the same methodology and template set as described in \citet{finkelstein24} (including the updated templates from \citealt{larson23b}, and excluding the \textit{HST}/WFC3 photometry).

\subsection{Spectroscopy}\label{sec:spectroscopy}
We analyze data from the Cycle 1 CEERS survey \citep[\textit{JWST} ERS\#1345, PI Finkelstein;][]{finkelstein17,finkelstein25} and the Cycle 2 RUBIES program \citep[\textit{JWST} Cycle 2 GO\#4233, PIs de Graaf and Brammer;][]{degraaff23,degraaff24}.

In brief, CEERS consists of eight NIRSpec MSA pointings in the EGS field (hereafter referred to as the CEERS field). Six of these pointings have data from four NIRSpec configurations: PRISM/CLEAR\footnote{Note: two of these PRISM pointings were corrupted by an MSA short circuit, and the individual PRISM spectra from these pointings were visually inspected for corruption before use in this analysis.} G140M/F100LP, G235M/F170LP, and G395M/F290LP. Two additional pointings only have PRISM data.  Each configuration and each pointing used three one-shutter nods and NRSIRS2 readout mode with 14 groups per nod (with a single integration per nod) for a total of 3107 seconds of exposure time. 

RUBIES consists of six pointings in the CEERS (EGS) field, and 12 pointings in the PRIMER-UDS field. All of these pointings have PRISM and G395M configuration data, although not every object observed has data in both configurations. Each configuration and each pointing used three one-shutter nods and NRSIRS2RAPID readout mode with 65 groups per nod (with a single integration per nod) for a total of 2889 seconds of exposure time.

As there are no G140M or G235M RUBIES data, and to best manage the scope of this investigation, we limit our search for BLAGN to the combined samples of G395M data from both programs and supplement with PRISM data for additional measurements when available.
Throughout this work, we construct object names as the combination of the observing program and field (CEERS, RUBIES-EGS, and RUBIES-UDS) and the source's MSA ID, separated by an hyphen. In total, we examine 677 G395M spectra from CEERS and 4521 G395M spectra from RUBIES. 

\subsubsection{Data Reduction}\label{sec:reduction}
In order to take advantage of the continuous updates to the \textit{JWST} Data Calibration Pipeline, we reprocess the CEERS and RUBIES spectroscopic data using pipeline version 1.13.4 (DMS Build B10.1) and CRDS version 1215.pmap. For each program, we download the uncalibrated data (\texttt{uncal} files) and NIRSpec MSA Metafiles from MAST as our starting point.

In Stage 1 of the pipeline, we use the default parameters specified by CRDS 1215.pmap, for all steps except the \texttt{jump} step, where we specify the following parameters to best reject ``snowball'' artifacts: \texttt{expand\_large\_events}:~True, \texttt{after\_jump\_flag\_dn1}:~0, \texttt{after\_jump\_flag\_time1}:~0, \texttt{after\_jump\_flag\_dn2}:~0, \texttt{after\_jump\_flag\_time2}:~0, \texttt{min\_sat\_area}:~15.0, and \texttt{expand\_factor}:~2.0, as in \cite{arrabalharo23}. In Stages 2 and 3, we keep all parameters set to the defaults specified by the CRDS, including keeping the default pathloss correction enabled. 

\subsubsection{Flux Extraction and Noise Calibration}\label{sec:extraction}
To best optimize the extraction of 1D spectra from the \textit{JWST} Pipeline Level 3 \texttt{s2d} 2D spectra data products, we adopt the methodology of \cite{horne86}. In short, for a given spectrum, we first construct a spatial profile by taking the median of the 2D spectrum along the spectral direction. We next isolate the source's central trace by setting all pixels in the spatial profile that are not a part of the center-most positive feature to zero. We then normalize the area under this masked spatial profile to one. Finally, we follow the prescription given in Table~1 of \cite{horne86} using this normalized profile as the variable P. This prescription produces an optimally extracted 1D spectrum and error spectrum designed to maximize the spectral signal to noise. We use these optimally extracted spectra for all of the data analyses in this work.

In previous NIRSpec studies \citep[e.g.][]{maseda23}, the errors computed by the \textit{JWST} pipeline were frequently reported to be underestimated. To test for this effect in our reprocessed data, we compare the error arrays for each G395M spectrum to the standard deviation of the science spectrum, calculated as follows. Separately from our line fitting (described below), we first remove the overall continuum shape by smoothing a spectrum using a 101-element wide median filter. We subtract this smoothed spectrum from the original to remove the continuum shape. We then use sigma-clipped statistics to determine the standard deviation of the flux residuals in the spectrum while removing areas of corruption, artifacts, and emission lines. When we compare the median of the spectrum's pipeline-derived error array to this empirical sigma-clipped standard deviation, we find that the empirical error is $67^{+20}_{-22}\%$ larger than the error arrays for the median spectrum in our datasets. \cite{maseda23} also reported a similar value of 70\% for G395M MSA data in their NIRSpec study (using pipeline version 1.10.2 and CRDS 1097.pmap). We therefore re-scale all of our G395M error arrays by the more conservative factor of 1.7 to better represent the true empirically measured uncertainties in the data. 

\section{Data Analysis}\label{sec:analysis}

\subsection{H$\alpha$ Line Detection and Fitting}\label{sec:linedetection}
We search for \Ha broad-lines in every CEERS and RUBIES G395M spectrum in a multiple-step process inspired in part by the methodology outlined in \cite{larson23}. Here, we first describe our emission-line discovery method, and discuss below how we identify lines as H$\alpha$.

Initially, we perform a pre-selection to identify candidate lines of interest (at low computational cost). In this pre-selection, we remove the overall continuum shape by smoothing the spectrum using a 151-element wide median filter ($>15,000$~km~s\per in width, such that even the broadest of emission lines will be unaffected). We subtract this smoothed spectrum from the original to remove the continuum shape. We then examine every pixel in a spectrum and select pixels with a signal-to-noise ratio (S/N) greater than five to search for the peaks of potential emission lines. At each of these pixels, we fit a model broad-line with both narrow and broad components centered on that pixel (in a non-continuum subtracted spectrum) using the \texttt{curve\_fit} routine from the \texttt{scipy} package for Python \citep{scipy20}. This model consists of the sum of two degenerate Gaussians that share a common central wavelength with independent amplitudes and full-width-half-maxima (FWHMs), and a linear model to account for any underlying (typically very weak or non-detected in G395M) continuum. In all of our models, we forward-model the effects of instrumental broadening by assuming a diffraction-limited point source and adding the modeled line width and the spectral resolution of the G395M grating ($\sim$230~km~s$^{-1}$) in quadrature such that we may fit the underlying kinematic FWHM of measured lines. As such, hereafter all given FWHM values are inherently corrected for the effects of instrumental broadening. In these fits we use flat uninformative priors for all model parameters and require that the FWHM and amplitude of the two model Gaussians be greater than zero, and that the shared line center be within two pixels ($400-700$~km~s\per, depending on the wavelength) of the detected SNR$>$5 pixel. 

We next examine the fitted broad-line model and enforce several criteria to select high-fidelity broad emission lines.  First, we require the peak of the combined model (the sum of the peaks of the broad and narrow components, and the continuum offset) is at least 80\% of the peak observed flux in the fitted region. We then require that the broader of the two fitted Gaussians has a FWHM greater than 500~km~s$^{-1}$. In later steps, we further restrict this value to $>$700~km~s$^{-1}$ and introduce additional components to model the [N\,\textsc{ii}] doublet (assuming that the broad-line is \Ha), but we initially leave this cut more conservative so as to avoid removing potential BLAGN objects before conducting more robust Bayesian fitting. Finally, we require that the broader Gaussian have an amplitude that is at least 50\% of the median of the noise array in the fitted region. As before, we deliberately leave this cut quite conservative to avoid removing real BLAGN before our Bayesian fitting. These requirements ensure that a candidate pixel may indeed be the peak of a line with a significant broad component that warrants additional study and more sophisticated and computationally expensive fitting. Finally, we require that the pixel be far enough from the end of the spectral wavelength range such that the fitted broad-line's flux falls to 10\% of the peak line flux before the spectrum ends. This requirement helps to reject candidate lines that are simply the result of noisy data near the ends of the spectral wavelength range. After applying the above methodology to all of the CEERS and RUBIES G395M spectra, we discover a population of 1366 candidate broad-lines out of a total of 5198 spectra. 

\begin{figure}[t]
\centering
\includegraphics[angle=0,width=\columnwidth]{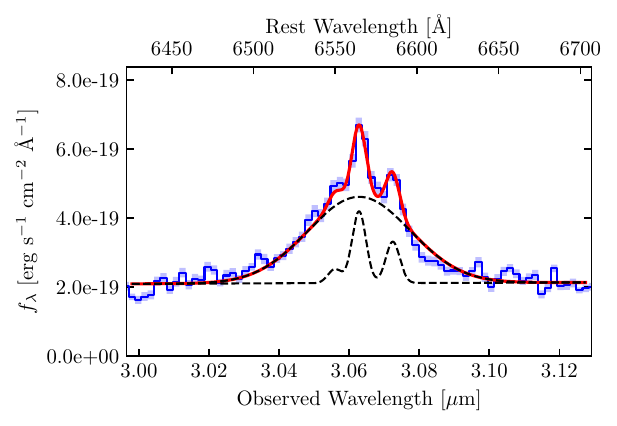}
\caption{\Ha and [N\,\textsc{II}]$\lambda\lambda$6548,6583 line fits to the G395M spectrum for object RUBIES-EGS-15825, a confirmed $z=3.67$ broad-line object in our sample. Here, the spectrum is shown in blue with 1$\sigma$ uncertainties shown as a shaded blue region, the broad \Ha component and the narrow \Ha and [N\,\textsc{II}]$\lambda\lambda$6548,6583 components of the fitted model are shown as black dashed curves, and the combined model is shown as a red curve. This model accounts for all of the expected features in the \Ha region emitted by a BLAGN.}
\label{fig:Ha_fit}
\end{figure}

We next re-fit these 1366 candidate broad-lines using Markov-Chain-Monte-Carlo (MCMC) fitting using the \texttt{emcee} package for \texttt{Python} \citep{emcee}. Here we model the \Ha and [N\textsc{ii}] lines using a five-component model. We model a \Ha line with a narrow Gaussian component and a broad Gaussian component that share a common line center. We also model [N\,\textsc{II}]$\lambda\lambda$6548,6583 as two Gaussians with a fixed internal amplitude ratio of 1:2.94 \citep{osterbrock06}, a common line-width with the narrower \Ha component, and line centers that assume the same redshift as the modeled \Ha line. Finally, we include a simple linear model to account for any faintly detected continuum emission. We show an example BLAGN object (RUBIES-EGS-15825) that exhibits significant [N\,\textsc{II}]$\lambda\lambda$6548,6583 emission to best illustrate our BLAGN model in Figure~\ref{fig:Ha_fit}. For this model, we impose a number of priors such that we require: (1) positive line fluxes for all line components, (2) the narrow component of \Ha (and [N\textsc{ii}]) is narrower than the broad component and narrower than 500~km~s\per, (3) the broad component of \Ha has a FWHM  $<$10,000~km~s$^{-1}$, (4) the \Ha line center is within 250~km~s$^{-1}$ of the previously fitted line center, and (5) the continuum offset is within the 10th and 90th percentile of the flux values in the fitted spectral region. We fit the portion of the full object spectrum that lies within 6500~km~s$^{-1}$ of the previously fitted line center. We initialize 32 MCMC ``walkers'' using the \texttt{curve\_fit} best-fit parameters as starting points (after perturbing these values for each walker to ensure that the walkers are independent) and allow them to explore the parameter space for 5000 steps per walker. To minimize the effects of chain ``burn-in'' we allow \texttt{emcee} to discard the first 2/3 of the chain, and use the last 1667 steps from each walker for our analysis, resulting in 53344 element posterior distributions for each fitted parameter. During this process, we save the full chains for each fitted parameter so that we may easily calculate the distributions of and thus the uncertainties on derived parameters in later steps in our analysis. 

We next require that the MCMC fitted broad components of \Ha have $\textrm{FWHM}>700$~km~s$^{-1}$. We choose this value to provide a balance between probing low-mass AGN while not contaminating our sample with non-AGN objects. We later compensate for potentially missed low-mass AGN in our completeness corrections (\S\ref{sec:completeness}).  This cut reduces the sample from 1366 candidates, to 1142 candidates. We next require that the broad components be detected with a flux S/N$>4$, reducing the sample from 1142 to 176 candidates. This drastic reduction in candidate sample size is by-design, as our pre-selection cuts are deliberately weak. Of the 1142 candidates, 685 exhibit a broad component with S/N$<$2, 215 exhibit a broad component with 2$<$S/N$<$3, and 66 exhibit a broad component with 3$<$S/N$<$4. It is unlikely that the S/N$<$3 objects show real broad-line components, and likely only sample noise. While a fraction of the 3$<$S/N$<$4 objects may be BLAGN, but we compensate for these objects in our completeness correction (see \S\ref{sec:completeness}). These requirements produce a sample of 176 spectra. 

We next re-fit fit each object spectrum with a modified version of our above model that excludes the broad H$\alpha$ line component as a standard of comparison to test for the significance of the broad line. We compute the Bayesian Information Criterion (BIC) for each model fit. We next use the difference in the BIC between the models to determine which model fits best for each candidate line (with the lower BIC model preferred). We require a $\Delta$BIC~$>6$ in favor of the broad-line model over the narrowline model, suggesting that the inclusion of the broad-line significantly improves the overall spectral fit without overfitting or adding unnecessary components to the model. This requirement reduces the sample of 176 objects to 162 objects, implying that the previously fitted broad-line components in the rejected 14 objects were not significant.

Finally, we require that all spectra pass a visual inspection to reject artifacts and contaminants, and to verify the objects have clear and unambiguous redshifts. For this, we require that the redshift be based on the presence of at least three strong emission lines (with one typically being \Ha and the others being the [O\,\textsc{iii}]$\lambda\lambda$4959,5007 doublet, the [S\,\textsc{ii}]$\lambda\lambda$6718,6733 doublet, and/or the [S\,\textsc{iii}]$\lambda\lambda$9069,9531 doublet) in the grating or prism spectrum (if available) of an object. We find secure redshifts and observe clean visual inspections for 79 of the 162 remaining objects, with most of the removed candidates ($\sim75\%$) clearly being contamination in the 2D spectra from either artifacts or overlapping spectra, with some others ($\sim15\%$) showing only a single apparent emission line in the G395M spectrum that cannot be used to determine a robust multi-line redshift solution. Finally, the last $\sim10\%$ of the rejected spectra feature blended lines (such as the [S\,\textsc{ii}]$\lambda\lambda6718,6733$ doublet) that were fit as broad-lines. We stress that we do not evaluate the significance of the the broad-line using this visual inspection, rather we only reject objects that are clearly artifacts or mis-fit doublets.

While our broad-line search is designed principally to detect \Ha emitters, our methodology is also capable of selecting other broad-lines. In these cases, the amplitude of the modeled [N\,\textsc{ii}] doublet is simply minimized during the fitting process. As such, of the 79 broad-line spectra with secure redshifts, 63 spectra feature a broad \Ha line. The other 16 spectra that do not contain broad \Ha consist of one Pa-$\gamma$ line (first identified in \citealt{wang24a}), seven He\,\textsc{i}~$\lambda10830$ lines, three [S\,\textsc{iii}]~$\lambda9531$ lines, and five Pa-$\alpha$ lines. While these $z<3.5$ objects are interesting in their own right, here we limit our study to the sample of 63 broad-line \Ha spectra at $3.5<z<6.8$. 

\begin{figure}[t]
\centering
\includegraphics[angle=0,width=\columnwidth]{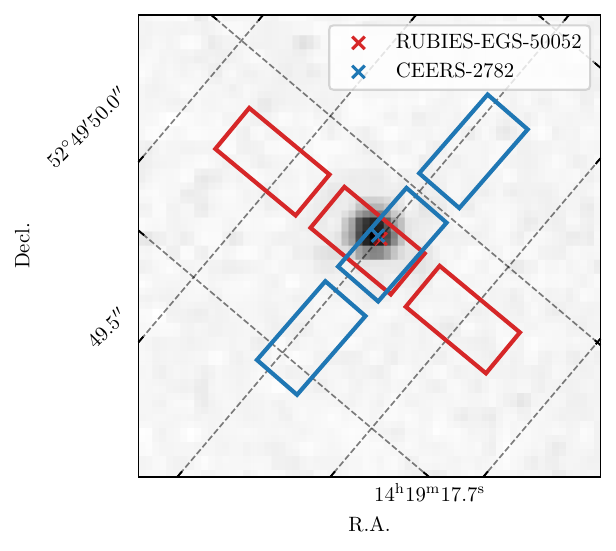}
\caption{2'' $\times$ 2'' NIRCam F444W (containing the \Ha line) cutout of CEERS-2782/RUBIES-EGS-50052. The blue/red outlines show the positions of the NIRSpec slitlets (0\farcs 20 $\times$ 0\farcs 46 per shutter) in each observation. The RUBIES-EGS-50052 slitlet alignment is significantly better centered than the CEERS-2782 pointing. Due to this superior centering, we use the higher S/N RUBIES-EGS-50052 spectrum in this analysis.}
\label{fig:doublesource}
\end{figure}

One object in our sample was observed by both CEERS and RUBIES and was thus detected as both objects CEERS-2782 and RUBIES-EGS-50052 (and is in fact one of the two BLAGN first discovered in CEERS by \citealt{kocevski23}). Interestingly, the CEERS spectrum (CEERS-2782) only exhibits $\sim$35\% of the \Ha line flux observed in the RUBIES observation (before applying a flux correction, see \S\ref{sec:fluxcal}). We attribute this difference in observed flux primarily to severe slit-losses, as the CEERS observation slitlet was quite off-center (see Figure~\ref{fig:doublesource}). Due to the much higher S/N in the RUBIES-EGS-50052 observation, we use this observation for our analysis of this object. As such, we report a final sample of 62 unique broad-line objects. 

\subsection{Flux Calibration}\label{sec:fluxcal}
While NIRSpec MSA flux calibration has improved significantly since the first data releases and pipelines \citep[e.g.][]{schaerer22,taylor22,carnall23,trump23}, additional corrections to account for slit-losses/path-losses and overall photometric calibrations are still necessary. We use the NIRCam photometry described in \S\ref{sec:photometry} to account for these effects in our dataset. To do this, we first visibly inspect the 2D spectrum of each object for contamination. The spectra of some objects are contaminated by either positive flux from other sources in the same MSA row, or negative flux resulting from nodded background subtraction of sources in adjacent MSA rows. We visually inspect our sources and flag those that are impacted by either type of contamination. We find 11 sources with significant contamination, and 52 sources that are contamination free.

For contamination-free sources, we apply a simple flux correction based on NIRCam photometry. For each source, we use the NIRCam filter transmission curves and G395M spectrum to measure synthetic photometry from the G395M spectrum for both the F356W and F444W NIRCam filters. We then select the filter that overlaps the detected broad-line and scale the flux and error spectra such that its broadband spectroscopic flux density matches the flux density of the broadband photometry. While using the higher S/N PRISM spectra (when available) would be preferable for this calibration, due to the different MSA configurations between the PRISM and G395M data, the measured fluxes between the two are not directly comparable. For the remaining 11 contaminated sources and three sources that fall outside the CEERS NIRCam footprint (see Figure~\ref{fig:fields}), we instead apply the median corrective factor (1.28) from the 52 uncontaminated sources with NIRCam coverage to avoid any extreme calibration errors from the contamination. 

After this correction, we re-fit the broad-lines using the \texttt{emcee} and above-mentioned broad-line model to produce final, flux-calibrated line fits and calibrated line measurements using the full calibrated line fit posterior distributions. We note that four of our objects (RUBIES-EGS-28812, RUBIES-EGS-42046, RUBIES-EGS-49140, and RUBIES-UDS-146995) visually show significant narrow absorption features on their \Ha lines, similar to those exhibited in sources identified in \cite{matthee24} and \cite{kocevski24}. When refitting these objects, we introduce a additional absorption component to our line model. We model this component as a Gaussian with a strictly negative amplitude and allow its line center and FWHM to vary freely. Notably, we observe blue-ward velocity offsets of the absorption features relative to the center of the \Ha emission lines of $32^{+22}_{-23}$~km~s$^{-1}$, $172^{+30}_{-32}$~km~s$^{-1}$, $184^{+24}_{-14}$~km~s$^{-1}$, and $50^{+31}_{-15}$~km~s$^{-1}$, for RUBIES-UDS-146995, RUBIES-EGS-28812, RUBIES-EGS-42046, and RUBIES-EGS-49140, respectively. While the physical interpretation of these features is actively discussed in the literature \citep[e.g.,][]{matthee24,inayoshi24}, we defer a full analysis of these absorption features to future work.

We apply a similar procedure to calibrate the PRISM spectra (when available) of the broad-line sample. Here, for a given source, we use the transmission curves of the F150W, F200W, F277W, F365W, F410M, and F444W filters to measure broadband fluxes for each filter. We then fit a first order polynomial to the ratio of the fluxes measured from the spectra and the fluxes reported in the broadband photometry as a function of the logarithm of the filter centers. We use this best fit line as a wavelength dependent multiplicative correction curve, and apply this correction to the spectral data. We repeat this process five times to ensure reasonable convergence between the spectrum and measured photometry. We use these calibrated PRISM spectra in \S\ref{sec:lrds} to measure optical continuum slopes.

\subsection{Outflows}\label{sec:outflows}

We next test our sample of broad-line objects for evidence of inflows/outflows to help distinguish true BLAGN from objects that have broad emission components due to strong kinematics in the ISM. For the 25 objects at redshifts $z>5.5$, such that the [O\,\textsc{iii}]$\lambda\lambda$4959,5007 doublet falls within the G395M spectra, we examine the [O\,\textsc{iii}]$\lambda\lambda$4959,5007 doublet for evidence of broad lines. 

Here, we design two models for MCMC fitting. In the first model (the outflow model), we model [O\,\textsc{iii}]$\lambda\lambda$4959,5007 each as two Gaussians (one narrow and one broad). We enforce a common narrow and a common broad FWHM for both lines in the [O\,\textsc{iii}]$\lambda\lambda$4959,5007 doublet. In contrast to our H$\alpha$ broad-line search, here we allow the line centers of the narrow and broad line components to vary to account for a velocity offset from an outflow. We fix the amplitude ratio of the [O\,\textsc{iii}] doublet lines to 1:2.98 for both the narrow and broad components. We also include a simple linear model for any faint continuum. In the second model (the BLAGN model), we exclude the broad components of the [O\,\textsc{iii}]$\lambda\lambda$4959,5007. 

\begin{figure}[t]
\centering
\includegraphics[angle=0,width=\columnwidth]{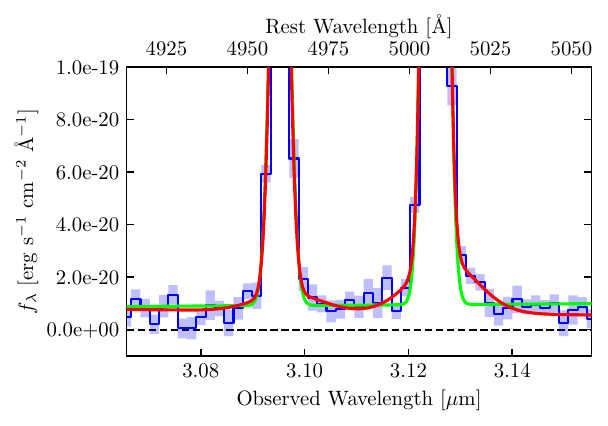}
\caption{Broad+narrow and single component fits to the [O\,\textsc{iii}]$\lambda\lambda$4959,5007 doublet in object RUBIES-EGS-50052. Here, the two component fit is shown by the red curve, the single component fits is shown by the green curve, and the spectrum is shown in blue. The double component fit significantly improves the line fit near the wings of the [O\,\textsc{iii}]$\lambda\lambda$4959,5007 lines, indicating the presence of an outflow.}
\label{fig:outflow}
\end{figure}

We fit the spectra with both the outflow and BLAGN models each using 32 walkers and a 5000 step chain, of which we discard the first 3333 steps. Here, we also compute the Bayesian Information Criterion (BIC) for each model fit: $\textrm{BIC}=k\ln(n)-2\ln(\hat{L})$, where $k$ is the number of free parameters in a fitted model, $\hat{L}$ is the maximized likelihood function, and $n$ is the number of fitted data points. We find only one object (RUBIES-EGS-50052) with BIC$_{outflow}$-BIC$_{BLAGN}<-6$ and a broad-line SNR$>4$, indicating significant broad [O\,\textsc{iii}]$\lambda\lambda$4959,5007 emission likely caused by a strong outflow. We show the single and double component fits to this object in Figure~\ref{fig:outflow}. This object exhibits an [O\,\textsc{iii}]$\lambda\lambda$4959,5007 broad component with a FWHM of $1409^{+553}_{-353}$~km~s$^{-1}$, and an \Ha broad component with a FWHM of $2131^{+88}_{-79}$~km~s$^{-1}$. These FWHMs demonstrate the [O\,\textsc{iii}]$\lambda\lambda$4959,5007 doublet has a slightly lower velocity broad component than H$\alpha$, possibly due to kinematics independent of the BLAGN. Curiously, RUBIES-EGS-50052 is the source that was observed by both RUBIES and CEERS (as CEERS-2782). Its CEERS spectrum shows no trace of an outflow, but due to the aforementioned expected slit-losses for this source in CEERS, we expect that the absence of detected outflows may simply be due to the lower S/N of the object in the CEERS data relative to the RUBIES data. Alternatively, the outflow component may be spatially dependent, and may be better sampled by the RUBIES shutter than the CEERS shutter. While the exact nature of these spatially dependent effects is beyond the scope of this work, the difference in the FWHMs of the broad \Ha emission and broad [O\,\textsc{iii}]emission from this source suggest a BLAGN signature, and we therefore keep RUBIES-EGS-50052 in our BLAGN sample. Nonetheless, if the broad \Ha emission from RUBIES-EGS-50052 is indeed the result of an outflow and not an AGN, our large 62 object sample size ensures that the inclusion or exclusion of RUBIES-EGS-50052 will not significantly affect our results.

Though we cannot perform this analysis for $z < 5.5$ objects in our sample, the 24/25 objects that show clear BLAGN signatures that kinematically dominate any outflow velocity structure suggest that our full sample is likely significantly dominated by BLAGN, with low likelihood of containing a significant number of outflow dominated objects whose broad \Ha emission is not driven by a central AGN. This result is consistent with the analysis performed in other collections of broad-line AGN detected by \textit{JWST} at $z>4$. Notably, \cite{Maiolino_2024_Xray} also finds that outflows play a negligible role in broadening the lines of the Type-I AGN used in their analysis.

We also repeat our outflow test on a stack of all 25 objects. Here, we stack these object spectra using an inverse variance weighting. Our outflow test indicates that an [O\,\textsc{iii}]$\lambda\lambda$4959,5007 broad-component is very weakly preferred in the stacked spectrum with a $\Delta$BIC$=4.26$. However this broad component is only detected with a SNR=1.1, and is thus not significant. As such, we maintain our conclusion that our sample indeed contains BLAGN and is not significantly contaminated by objects with strong outflows.

Beyond outflows, \cite{Baggen24} propose that the broad-lines seen in high-redshift galaxies with \textit{JWST} may instead be generated by highly-compact galaxies with intrinsically high virial velocities that can mimic the broad-lines typically associated with BLAGN. However, as 59/62 objects in our sample show both broad and narrow components in \Ha, we expect that these are indeed BLAGN showing narrow emission from galaxy kinematics (or the AGN narrow line regions) and broad emission from the AGN, and are not the exotically dense galaxies proposed by \cite{Baggen24}. While the final object (RUBIES-EGS-49140) is analyzed in \cite{wang24b} and studied in \cite{Baggen24} as a candidate for non-AGN broad-lines, we nonetheless keep this object in our BLAGN sample as the \cite{Baggen24} model does not preclude this object from being a BLAGN, it simply offers a novel alternate explanation for its broad spectral features. 

\vskip -24pt
\startlongtable
\centerwidetable
\begin{deluxetable*}{ccccccccc}
\tablecaption{Sample of BLAGN \label{tab:sample}}
\tablehead{
Source ID & R.A. & Decl. & $z_{spec}$ & $F_{H \alpha}$ & $F_{H \alpha, narrow}$ & $F_{H \alpha, broad}$ & FWHM$_{H \alpha, broad}$ & $\log_{10}(M_{BH})$ \\
 & [deg] & [deg] &  & \multicolumn{3}{c}{[$10^{-18}$ erg s$^{-1}$ cm$^{-2}$]} & [km s$^{-1}$] & $[M_{\odot}]$}
 \startdata
RUBIES-UDS-44043\textsuperscript{\textdagger} & 34.241817 & -5.239436 & 3.499 & 31.53$^{+1.66}_{-1.49}$ & 8.33$^{+1.09}_{-1.10}$ & 23.25$^{+1.70}_{-1.63}$ & 2728$^{+240}_{-208}$ & 7.66$^{+0.08}_{-0.07}$ \\
RUBIES-EGS-50812 & 214.845487 & 52.848281 & 3.519 & 12.07$^{+0.60}_{-0.58}$ & 5.85$^{+0.57}_{-0.58}$ & 6.24$^{+0.56}_{-0.53}$ & 1358$^{+224}_{-180}$ & 6.77$^{+0.14}_{-0.13}$ \\
RUBIES-UDS-21944 & 34.469218 & -5.283563 & 3.526 & 13.69$^{+1.26}_{-1.26}$ & 2.14$^{+0.71}_{-0.64}$ & 11.54$^{+1.46}_{-1.53}$ & 1459$^{+162}_{-175}$ & 6.96$^{+0.11}_{-0.13}$ \\
RUBIES-UDS-46566 & 34.293543 & -5.234588 & 3.538 & 124.88$^{+2.01}_{-2.11}$ & 74.50$^{+6.51}_{-6.48}$ & 50.39$^{+6.34}_{-6.41}$ & 747$^{+56}_{-51}$ & 6.67$^{+0.05}_{-0.05}$ \\
RUBIES-UDS-154183 & 34.410749 & -5.129664 & 3.546 & 58.67$^{+1.05}_{-1.02}$ & 9.25$^{+0.90}_{-0.95}$ & 49.45$^{+1.23}_{-1.20}$ & 2234$^{+68}_{-69}$ & 7.64$^{+0.03}_{-0.03}$ \\
RUBIES-UDS-153207 & 34.493112 & -5.130999 & 3.597 & 310.80$^{+6.08}_{-5.80}$ & 127.70$^{+25.46}_{-29.04}$ & 183.25$^{+26.44}_{-22.74}$ & 857$^{+82}_{-67}$ & 7.06$^{+0.06}_{-0.05}$ \\
RUBIES-EGS-50522 & 214.855980 & 52.854661 & 3.614 & 4.61$^{+0.74}_{-0.61}$ & 1.34$^{+0.41}_{-0.35}$ & 3.28$^{+0.84}_{-0.82}$ & 1733$^{+297}_{-317}$ & 6.87$^{+0.17}_{-0.21}$ \\
RUBIES-EGS-920 & 215.052344 & 52.884268 & 3.616 & 34.24$^{+0.98}_{-0.84}$ & 28.12$^{+1.83}_{-2.50}$ & 6.16$^{+2.11}_{-1.34}$ & 823$^{+451}_{-203}$ & 6.33$^{+0.36}_{-0.20}$ \\
RUBIES-UDS-150323 & 34.417822 & -5.134842 & 3.618 & 7.51$^{+0.26}_{-0.27}$ & 2.91$^{+0.21}_{-0.22}$ & 4.61$^{+0.26}_{-0.26}$ & 2178$^{+168}_{-157}$ & 7.15$^{+0.07}_{-0.07}$ \\
RUBIES-EGS-58237 & 214.850571 & 52.866030 & 3.651 & 63.78$^{+2.43}_{-2.38}$ & 14.34$^{+2.08}_{-2.36}$ & 49.47$^{+3.92}_{-3.90}$ & 2137$^{+121}_{-109}$ & 7.62$^{+0.04}_{-0.04}$ \\
RUBIES-UDS-61627 & 34.238394 & -5.205775 & 3.654 & 8.40$^{+0.52}_{-0.49}$ & 4.36$^{+0.47}_{-0.41}$ & 4.04$^{+0.65}_{-0.69}$ & 1515$^{+258}_{-217}$ & 6.80$^{+0.15}_{-0.15}$ \\
RUBIES-UDS-5496 & 34.405872 & -5.312951 & 3.655 & 13.30$^{+1.07}_{-0.94}$ & 6.42$^{+0.57}_{-0.59}$ & 6.94$^{+1.03}_{-1.05}$ & 1657$^{+320}_{-267}$ & 6.99$^{+0.17}_{-0.17}$ \\
RUBIES-EGS-15825 & 215.079264 & 52.934252 & 3.666 & 72.08$^{+2.00}_{-2.06}$ & 7.29$^{+1.40}_{-1.32}$ & 64.80$^{+2.27}_{-2.46}$ & 3343$^{+189}_{-169}$ & 8.08$^{+0.05}_{-0.05}$ \\
RUBIES-UDS-18302 & 34.233628 & -5.283850 & 3.698 & 21.30$^{+3.88}_{-3.19}$ & 6.32$^{+2.45}_{-3.12}$ & 15.38$^{+2.88}_{-2.54}$ & 1425$^{+621}_{-519}$ & 7.01$^{+0.36}_{-0.41}$ \\
RUBIES-UDS-46885 & 34.291666 & -5.233788 & 3.730 & 25.73$^{+0.44}_{-0.40}$ & 15.46$^{+1.26}_{-1.32}$ & 10.31$^{+1.26}_{-1.25}$ & 709$^{+52}_{-45}$ & 6.32$^{+0.05}_{-0.05}$ \\
RUBIES-UDS-146995\textsuperscript{\textdaggerdbl} & 34.331043 & -5.139963 & 3.732 & 26.26$^{+1.43}_{-1.58}$ & 0.90$^{+0.83}_{-0.58}$ & 25.23$^{+1.42}_{-1.62}$ & 1403$^{+85}_{-89}$ & 7.11$^{+0.05}_{-0.06}$ \\
RUBIES-EGS-34978 & 214.861690 & 52.818438 & 3.772 & 91.02$^{+1.30}_{-1.27}$ & 68.01$^{+1.53}_{-1.53}$ & 23.04$^{+1.71}_{-1.75}$ & 1485$^{+76}_{-77}$ & 7.15$^{+0.04}_{-0.05}$ \\
RUBIES-EGS-19174 & 214.860840 & 52.784773 & 3.774 & 72.89$^{+2.39}_{-2.30}$ & 55.44$^{+4.26}_{-4.98}$ & 17.62$^{+4.61}_{-4.10}$ & 878$^{+248}_{-167}$ & 6.62$^{+0.20}_{-0.16}$ \\
RUBIES-UDS-10036\textsuperscript{\textdagger} & 34.381671 & -5.303742 & 3.806 & 9.46$^{+0.40}_{-0.36}$ & 3.12$^{+0.32}_{-0.29}$ & 6.34$^{+0.38}_{-0.39}$ & 1234$^{+107}_{-91}$ & 6.73$^{+0.08}_{-0.07}$ \\
RUBIES-EGS-37032\textsuperscript{\textdagger} & 214.849388 & 52.811824 & 3.850 & 18.00$^{+1.60}_{-1.25}$ & 5.09$^{+1.02}_{-1.06}$ & 13.08$^{+1.13}_{-1.17}$ & 1960$^{+347}_{-256}$ & 7.29$^{+0.15}_{-0.13}$ \\
CEERS-11728 & 215.084870 & 52.970738 & 3.869 & 9.88$^{+0.49}_{-0.45}$ & 7.08$^{+0.52}_{-0.59}$ & 2.86$^{+0.59}_{-0.59}$ & 1191$^{+266}_{-237}$ & 6.53$^{+0.18}_{-0.19}$ \\
RUBIES-UDS-22304 & 34.399170 & -5.283007 & 3.906 & 24.14$^{+0.52}_{-0.52}$ & 19.88$^{+1.13}_{-1.64}$ & 4.27$^{+1.48}_{-0.95}$ & 844$^{+197}_{-153}$ & 6.32$^{+0.15}_{-0.13}$ \\
RUBIES-UDS-16053\textsuperscript{\textdagger} & 34.367104 & -5.293524 & 3.952 & 8.18$^{+0.64}_{-0.60}$ & 2.13$^{+0.38}_{-0.36}$ & 6.06$^{+0.63}_{-0.60}$ & 1198$^{+183}_{-154}$ & 6.71$^{+0.13}_{-0.13}$ \\
RUBIES-UDS-147411 & 34.360718 & -5.139081 & 3.966 & 8.42$^{+0.48}_{-0.46}$ & 3.33$^{+0.70}_{-0.65}$ & 5.11$^{+0.58}_{-0.65}$ & 1100$^{+225}_{-175}$ & 6.60$^{+0.16}_{-0.15}$ \\
RUBIES-UDS-11721 & 34.411039 & -5.300780 & 3.978 & 33.51$^{+1.71}_{-1.66}$ & 23.75$^{+1.42}_{-1.30}$ & 9.78$^{+2.25}_{-2.30}$ & 1937$^{+365}_{-293}$ & 7.24$^{+0.16}_{-0.16}$ \\
RUBIES-UDS-8895 & 34.363041 & -5.306108 & 3.982 & 120.33$^{+2.81}_{-2.93}$ & 67.12$^{+2.79}_{-2.75}$ & 53.25$^{+3.95}_{-4.29}$ & 1523$^{+71}_{-74}$ & 7.37$^{+0.05}_{-0.05}$ \\
RUBIES-UDS-30969 & 34.296356 & -5.268672 & 4.000 & 51.14$^{+0.79}_{-0.75}$ & 40.29$^{+1.34}_{-1.54}$ & 10.92$^{+1.44}_{-1.37}$ & 1079$^{+103}_{-107}$ & 6.74$^{+0.07}_{-0.08}$ \\
RUBIES-UDS-155916 & 34.317031 & -5.127611 & 4.098 & 7.93$^{+0.88}_{-0.95}$ & 0.16$^{+0.30}_{-0.12}$ & 7.72$^{+0.94}_{-1.09}$ & 1646$^{+190}_{-189}$ & 7.06$^{+0.11}_{-0.13}$ \\
RUBIES-UDS-31747\textsuperscript{\textdagger} & 34.223757 & -5.260245 & 4.130 & 16.16$^{+0.45}_{-0.51}$ & 4.85$^{+0.52}_{-0.58}$ & 11.32$^{+0.63}_{-0.70}$ & 1688$^{+93}_{-92}$ & 7.17$^{+0.05}_{-0.05}$ \\
RUBIES-UDS-119957\textsuperscript{\textdagger} & 34.268908 & -5.176722 & 4.149 & 12.83$^{+0.39}_{-0.39}$ & 4.23$^{+0.39}_{-0.36}$ & 8.61$^{+0.44}_{-0.48}$ & 1845$^{+121}_{-114}$ & 7.19$^{+0.06}_{-0.05}$ \\
RUBIES-EGS-28812\textsuperscript{\textdaggerdbl} & 214.924149 & 52.849050 & 4.222 & 17.59$^{+1.11}_{-0.96}$ & 7.24$^{+1.38}_{-1.22}$ & 10.37$^{+0.65}_{-0.65}$ & 2098$^{+120}_{-117}$ & 7.35$^{+0.04}_{-0.04}$ \\
RUBIES-UDS-143683 & 34.316389 & -5.144678 & 4.226 & 15.94$^{+0.56}_{-0.55}$ & 11.95$^{+0.51}_{-0.58}$ & 4.05$^{+0.47}_{-0.51}$ & 1426$^{+268}_{-271}$ & 6.81$^{+0.16}_{-0.19}$ \\
RUBIES-UDS-35974 & 34.331644 & -5.260593 & 4.367 & 41.29$^{+0.77}_{-0.76}$ & 33.18$^{+1.09}_{-1.16}$ & 8.16$^{+1.10}_{-1.11}$ & 1179$^{+133}_{-141}$ & 6.80$^{+0.09}_{-0.11}$ \\
RUBIES-UDS-63139 & 34.230848 & -5.202607 & 4.435 & 5.45$^{+0.26}_{-0.24}$ & 2.53$^{+0.27}_{-0.25}$ & 2.92$^{+0.34}_{-0.37}$ & 1107$^{+111}_{-96}$ & 6.55$^{+0.09}_{-0.09}$ \\
RUBIES-UDS-48507 & 34.284578 & -5.230702 & 4.468 & 19.68$^{+0.30}_{-0.30}$ & 11.22$^{+0.60}_{-0.63}$ & 8.47$^{+0.60}_{-0.62}$ & 762$^{+39}_{-37}$ & 6.43$^{+0.04}_{-0.04}$ \\
CEERS-1244 & 215.240652 & 53.036041 & 4.478 & 50.54$^{+0.83}_{-0.82}$ & 16.55$^{+0.71}_{-0.70}$ & 34.02$^{+0.98}_{-1.05}$ & 2619$^{+81}_{-73}$ & 7.82$^{+0.03}_{-0.02}$ \\
RUBIES-EGS-29489\textsuperscript{\textdagger} & 215.022071 & 52.920786 & 4.543 & 13.48$^{+0.74}_{-0.74}$ & 2.21$^{+0.63}_{-0.74}$ & 11.32$^{+0.78}_{-0.79}$ & 2055$^{+198}_{-185}$ & 7.39$^{+0.08}_{-0.09}$ \\
RUBIES-UDS-19484\textsuperscript{\textdagger} & 34.232426 & -5.280654 & 4.656 & 3.76$^{+0.32}_{-0.29}$ & 1.73$^{+0.36}_{-0.41}$ & 2.06$^{+0.39}_{-0.41}$ & 1278$^{+321}_{-288}$ & 6.62$^{+0.19}_{-0.22}$ \\
RUBIES-UDS-182791 & 34.213813 & -5.087050 & 4.718 & 40.12$^{+0.89}_{-0.84}$ & 8.20$^{+0.65}_{-0.61}$ & 31.91$^{+1.12}_{-1.08}$ & 3350$^{+132}_{-116}$ & 8.06$^{+0.03}_{-0.03}$ \\
RUBIES-EGS-45599 & 214.831103 & 52.824124 & 4.862 & 7.57$^{+0.35}_{-0.35}$ & 4.68$^{+0.29}_{-0.30}$ & 2.90$^{+0.35}_{-0.34}$ & 1498$^{+228}_{-203}$ & 6.86$^{+0.13}_{-0.13}$ \\
RUBIES-EGS-6411 & 215.109185 & 52.939770 & 4.880 & 7.26$^{+0.31}_{-0.26}$ & 4.96$^{+0.24}_{-0.23}$ & 2.32$^{+0.28}_{-0.27}$ & 1046$^{+191}_{-163}$ & 6.49$^{+0.16}_{-0.16}$ \\
RUBIES-EGS-42232\textsuperscript{\textdagger} & 214.886792 & 52.855381 & 4.954 & 23.59$^{+0.67}_{-0.68}$ & 4.48$^{+0.49}_{-0.53}$ & 19.14$^{+0.82}_{-0.86}$ & 2375$^{+96}_{-90}$ & 7.67$^{+0.04}_{-0.03}$ \\
RUBIES-EGS-46985 & 214.805654 & 52.809497 & 4.963 & 16.06$^{+1.15}_{-0.93}$ & 9.46$^{+0.53}_{-0.59}$ & 6.67$^{+0.77}_{-0.65}$ & 1290$^{+307}_{-247}$ & 6.90$^{+0.21}_{-0.20}$ \\
RUBIES-EGS-17146 & 214.949482 & 52.845415 & 5.000 & 36.80$^{+1.38}_{-1.29}$ & 21.82$^{+4.59}_{-5.21}$ & 15.01$^{+4.61}_{-3.90}$ & 763$^{+306}_{-150}$ & 6.60$^{+0.25}_{-0.15}$ \\
RUBIES-EGS-50052 & 214.823454 & 52.830277 & 5.240 & 37.85$^{+0.63}_{-0.66}$ & 20.07$^{+0.51}_{-0.51}$ & 17.78$^{+0.55}_{-0.56}$ & 2129$^{+82}_{-78}$ & 7.58$^{+0.03}_{-0.03}$ \\
CEERS-2782 & 214.823453 & 52.830281 & 5.240 & 26.41$^{+1.13}_{-1.18}$ & 14.47$^{+1.03}_{-1.03}$ & 11.95$^{+1.22}_{-1.19}$ & 1462$^{+239}_{-190}$ & 7.16$^{+0.14}_{-0.13}$ \\
RUBIES-EGS-13872 & 215.132933 & 52.970705 & 5.262 & 7.67$^{+0.48}_{-0.49}$ & 4.45$^{+0.44}_{-0.64}$ & 3.31$^{+0.45}_{-0.48}$ & 1223$^{+355}_{-345}$ & 6.74$^{+0.23}_{-0.28}$ \\
RUBIES-EGS-42046\textsuperscript{\textdagger}\textsuperscript{\textdaggerdbl} & 214.795368 & 52.788847 & 5.276 & 190.13$^{+4.61}_{-6.61}$ & 80.88$^{+4.61}_{-7.26}$ & 109.59$^{+2.10}_{-2.10}$ & 3146$^{+75}_{-67}$ & 8.30$^{+0.02}_{-0.02}$ \\
RUBIES-UDS-970351\textsuperscript{\textdagger} & 34.261900 & -5.105205 & 5.282 & 6.27$^{+0.73}_{-0.69}$ & 1.05$^{+0.38}_{-0.35}$ & 5.21$^{+0.73}_{-0.68}$ & 1657$^{+316}_{-265}$ & 7.11$^{+0.17}_{-0.17}$ \\
RUBIES-EGS-926125\textsuperscript{\textdagger} & 215.137081 & 52.988554 & 5.284 & 14.45$^{+0.41}_{-0.43}$ & 4.46$^{+0.33}_{-0.31}$ & 9.98$^{+0.46}_{-0.48}$ & 1888$^{+97}_{-92}$ & 7.36$^{+0.05}_{-0.05}$ \\
RUBIES-EGS-60935\textsuperscript{\textdagger} & 214.923373 & 52.925593 & 5.287 & 22.82$^{+0.56}_{-0.59}$ & 6.84$^{+0.56}_{-0.60}$ & 16.02$^{+0.64}_{-0.72}$ & 2144$^{+108}_{-108}$ & 7.57$^{+0.04}_{-0.04}$ \\
RUBIES-UDS-29813\textsuperscript{\textdagger} & 34.453355 & -5.270717 & 5.440 & 12.06$^{+0.55}_{-0.52}$ & 4.64$^{+0.35}_{-0.33}$ & 7.43$^{+0.55}_{-0.56}$ & 2047$^{+207}_{-188}$ & 7.39$^{+0.09}_{-0.09}$ \\
RUBIES-UDS-172350\textsuperscript{\textdagger} & 34.368951 & -5.103941 & 5.580 & 33.19$^{+0.84}_{-0.79}$ & 5.59$^{+0.68}_{-0.69}$ & 27.61$^{+0.92}_{-0.85}$ & 2068$^{+92}_{-78}$ & 7.68$^{+0.04}_{-0.04}$ \\
CEERS-746\textsuperscript{\textdagger} & 214.809142 & 52.868484 & 5.623 & 6.59$^{+0.53}_{-0.49}$ & 1.93$^{+0.46}_{-0.43}$ & 4.68$^{+0.55}_{-0.59}$ & 2021$^{+366}_{-289}$ & 7.29$^{+0.16}_{-0.14}$ \\
CEERS-672\textsuperscript{\textdagger} & 214.889677 & 52.832977 & 5.666 & 2.41$^{+0.39}_{-0.27}$ & 0.99$^{+0.19}_{-0.20}$ & 1.45$^{+0.32}_{-0.28}$ & 1433$^{+507}_{-345}$ & 6.75$^{+0.29}_{-0.26}$ \\
RUBIES-UDS-19521 & 34.383672 & -5.287732 & 5.669 & 11.29$^{+1.46}_{-1.38}$ & 5.46$^{+0.47}_{-0.49}$ & 5.84$^{+1.33}_{-1.18}$ & 1682$^{+400}_{-368}$ & 7.18$^{+0.22}_{-0.26}$ \\
RUBIES-UDS-47509\textsuperscript{\textdagger} & 34.264602 & -5.232586 & 5.673 & 22.92$^{+0.81}_{-0.83}$ & 9.05$^{+0.61}_{-0.62}$ & 13.87$^{+0.76}_{-0.80}$ & 2063$^{+157}_{-145}$ & 7.54$^{+0.07}_{-0.07}$ \\
RUBIES-EGS-37124\textsuperscript{\textdagger} & 214.990977 & 52.916524 & 5.682 & 9.94$^{+0.53}_{-0.55}$ & 1.93$^{+0.92}_{-0.74}$ & 7.98$^{+0.66}_{-0.73}$ & 1566$^{+246}_{-192}$ & 7.18$^{+0.12}_{-0.11}$ \\
RUBIES-UDS-139709\textsuperscript{\textdagger} & 34.296002 & -5.149895 & 5.685 & 39.47$^{+1.75}_{-1.66}$ & 11.11$^{+1.23}_{-1.15}$ & 28.37$^{+1.59}_{-1.55}$ & 2552$^{+205}_{-189}$ & 7.88$^{+0.07}_{-0.07}$ \\
CEERS-397 & 214.836197 & 52.882693 & 6.000 & 21.75$^{+2.40}_{-1.20}$ & 16.82$^{+0.44}_{-0.42}$ & 4.88$^{+2.48}_{-1.15}$ & 2727$^{+484}_{-413}$ & 7.60$^{+0.23}_{-0.18}$ \\
RUBIES-UDS-174752 & 34.205808 & -5.100500 & 6.039 & 14.82$^{+0.49}_{-0.50}$ & 8.54$^{+0.30}_{-0.30}$ & 6.28$^{+0.45}_{-0.44}$ & 1741$^{+140}_{-136}$ & 7.26$^{+0.08}_{-0.08}$ \\
RUBIES-EGS-49140\textsuperscript{\textdagger}\textsuperscript{\textdaggerdbl} & 214.892248 & 52.877410 & 6.685 & 116.81$^{+2.52}_{-3.23}$ & 4.94$^{+1.96}_{-2.27}$ & 111.71$^{+1.61}_{-1.65}$ & 2613$^{+48}_{-39}$ & 8.26$^{+0.02}_{-0.01}$ \\
RUBIES-UDS-807469 & 34.376139 & -5.310366 & 6.778 & 7.72$^{+0.48}_{-0.44}$ & 2.18$^{+0.46}_{-0.37}$ & 5.53$^{+0.37}_{-0.39}$ & 2044$^{+329}_{-218}$ & 7.43$^{+0.13}_{-0.11}$
\enddata
\tablecomments{The listed FWHM of the broad component is corrected for instrumental broadening. The stated errors on $M_{\textrm{BH}}$ do not take into account the systematic uncertainties from the \cite{reines13} prescription. We mark LRD sources with a (\textsuperscript{\textdagger}), and we mark the four sources fit with an absorption feature with a (\textsuperscript{\textdaggerdbl}). As noted in the text, RUBIES-EGS-50052 and CEERS-2782 are the same object observed independently by RUBIES and CEERS, and are thus marked with a (*). We attribute the differences in both line flux and FWHM to the extreme spatially dependent slit-loss seen in CEERS-2782.}
\end{deluxetable*}

\clearpage

\begin{figure*}[p]
\centering
\includegraphics[angle=0,width=\textwidth]{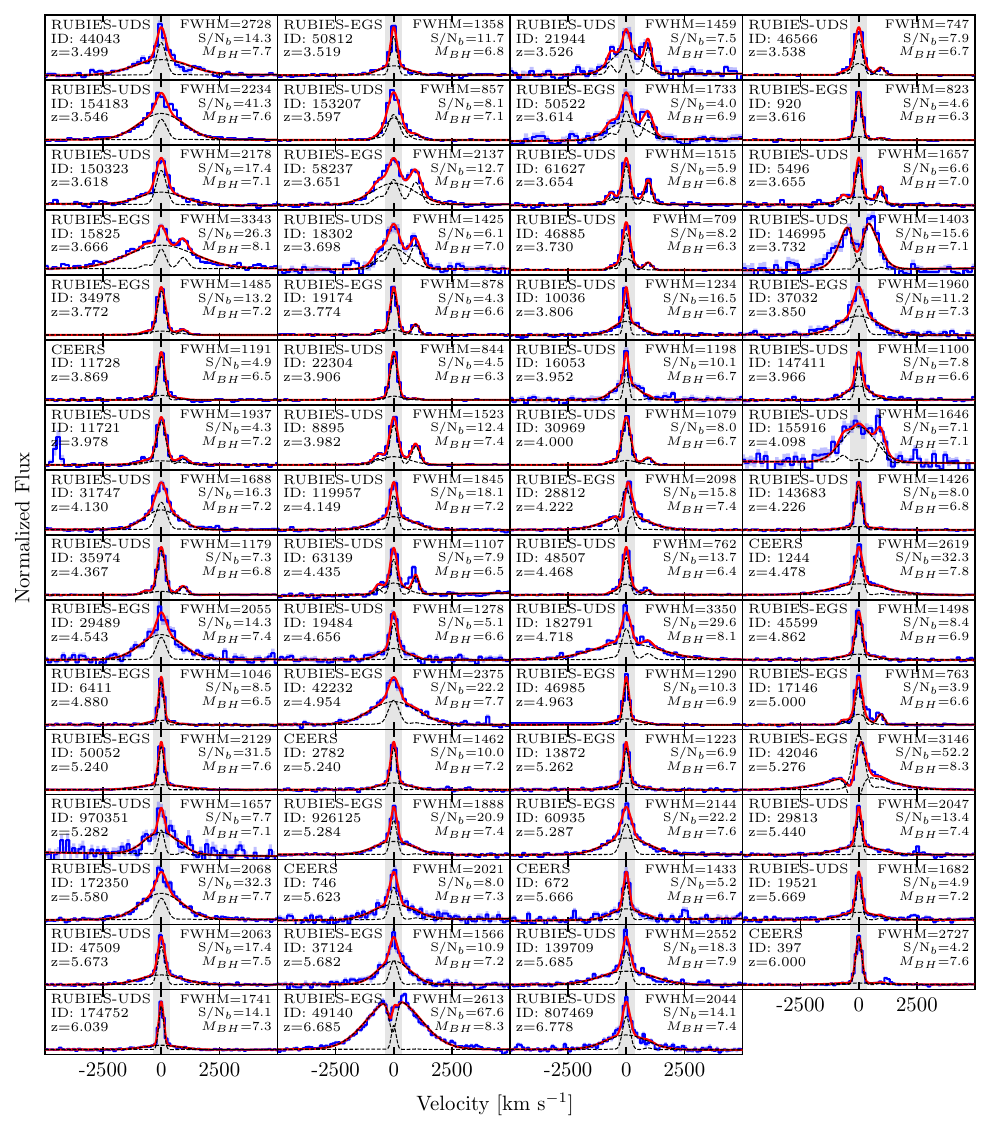}
\caption{The \Ha broad-lines (blue curves), broad-line fits (red curves), and broad and narrow components of the fits (black dashed curves) for all 63 spectra (for the 62 objects) in our BLAGN sample, sorted in order of increasing redshift. Here, we normalize each spectrum by the peak of the \Ha line fit, and plot the data and fits as a function of velocity relative to the \Ha line center. We show a 700~km~s\per region (our FWHM cut) as a shaded region centered at zero velocity in each panel. We give the MSA\_ID, redshift, broad component FWHM (in km~s\per), broad component signal-to-noise ratio, and black hole mass (in units of $\log_{10}\left(M_{\odot}\right)$) for each target in each panel. Note that CEERS-2782 and RUBIES-EGS-50052 are the same source observed by both CEERS and RUBIES.}
\label{fig:All_Broadlines}
\end{figure*}

\begin{figure*}[p]
\centering
\includegraphics[angle=0,width=\textwidth]{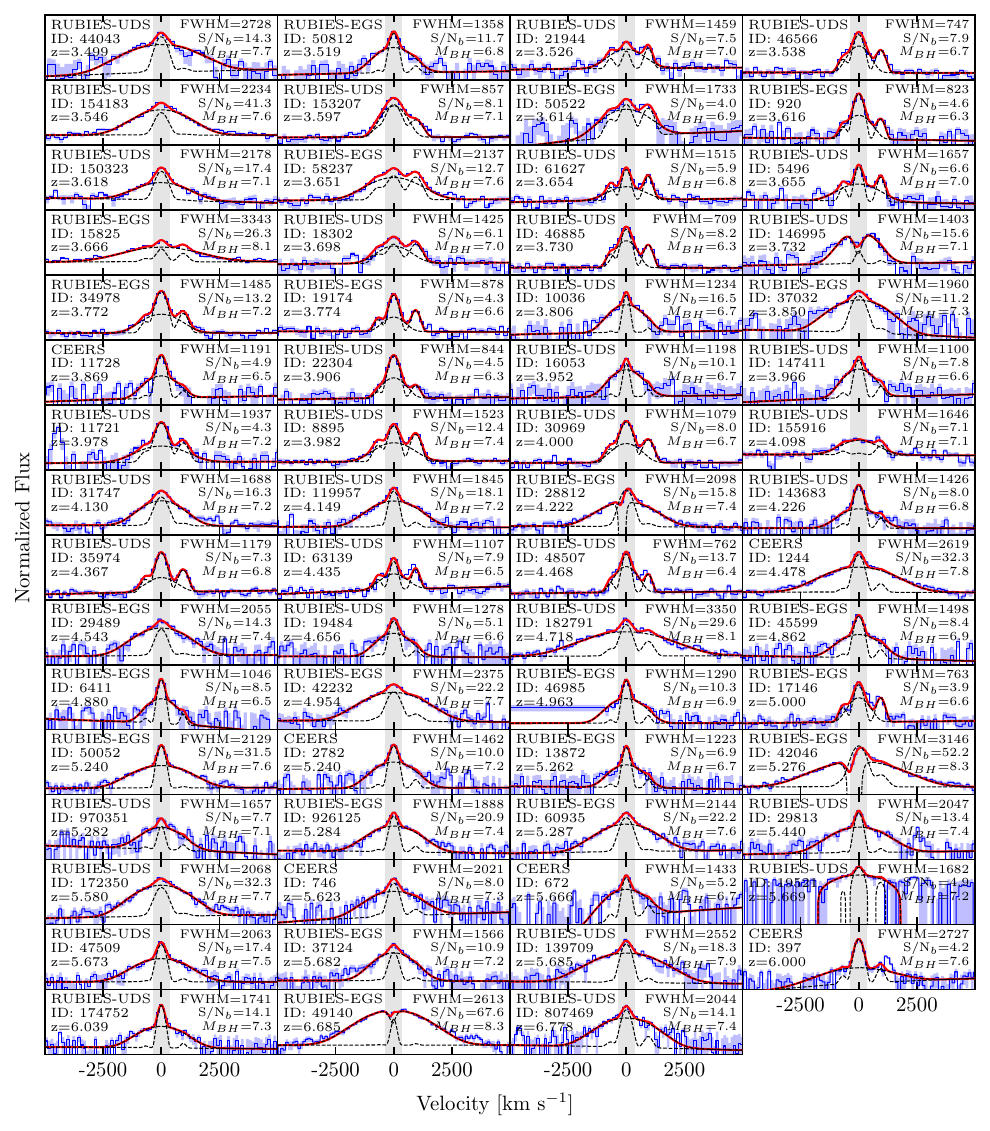}
\caption{The same as Figure~\ref{fig:All_Broadlines}, but here we plot the line fits and data on a logarithmic scale to best emphasize the broad components of each line. This helps in particular when the broad peak is much lower than the narrow peak, as can be seen in CEERS-397.}
\label{fig:All_Broadlines_Log}
\end{figure*}

\section{The BLAGN Sample}\label{sec:sample}
\begin{figure}[t]
\centering
\includegraphics[angle=0,width=\columnwidth]{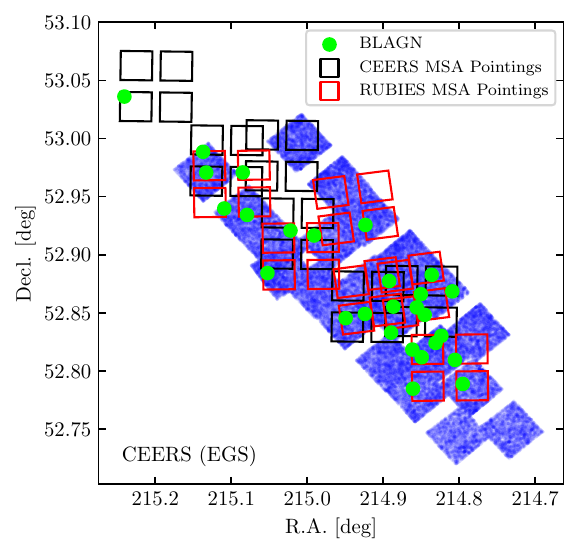}
\includegraphics[angle=0,width=\columnwidth]{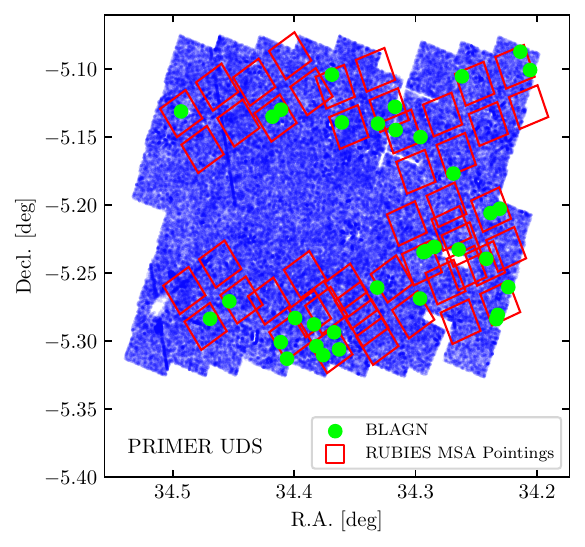}
\caption{Upper panel: the positions of the BLAGN sample (green circles), the CEERS NIRCam coverage (blue shading), and CEERS/RUBIES NIRSpec MSA footprints (black/red rectangles). Lower panel: the same as the top panel, for the PRIMER-UDS field. Note that the sky areas shown in the two panels are not exactly to scale with one another. The BLAGN are broadly distributed over the observed areas of the CEERS and PRIMER-UDS fields.}
\label{fig:fields}
\end{figure}

Here, we present our final sample of 62 \Ha spectroscopically confirmed BLAGN. We show galleries of the fitted \Ha broad-lines in Figure~\ref{fig:All_Broadlines} and Figure~\ref{fig:All_Broadlines_Log}, and detail the properties of these objects in Table~\ref{tab:sample}. We show the positions of the BLAGN in the CEERS (EGS) and PRIMER-UDS fields, as well as the footprints of the CEERS and RUBIES NIRSpec MSA pointings in Figure~\ref{fig:fields}. Three sources that lack NIRCam imaging coverage are visible in the upper left of the top panel and were targeted in the CEERS survey based on archival \textit{HST} CANDELS imaging \citep{grogin11,koekemoer11}. The sample clearly exhibits a large diversity in \Ha line shapes, [N\,\textsc{II}]$\lambda\lambda$6548,6583 line strengths, and redshifts. We analyze the properties of the sample below. 

\subsection{Comparison with Previous Samples}\label{sec:previoussamples}
We next compare our CEERS and RUBIES BLAGN samples to previously identified BLAGN samples in the literature. \cite{kocevski23} first identified two BLAGN that had been targeted for spectroscopic follow-up in CEERS after their photometric identification in previous studies (CEERS-2782: \citealt{labbe23,perezgonzalez23}, and CEERS-746: \citealt{onoue23}). \cite{harikane23} expanded the CEERS BLAGN sample to eight galaxies (including and recovering the \cite{kocevski23} sample). 

We detect seven of these eight sources in our sample (CEERS-397, CEERS-672, CEERS-746, CEERS-1244, CEERS-1665, and CEERS-2782), but fail to recover one object (CEERS-717) and reject two others (CEERS-1236, CEERS-1665). Upon examining CEERS-717, we find that the \cite{harikane23} custom data reduction extended the spectral coverage of the G395M to 5.3~$\mu$m, beyond the coverage returned by the standard pipeline (5.2~$\mu$m). As \cite{harikane23} report a spectroscopic redshift of $z=6.936$ (and thus a line center at 5.21~$\mu$m) for CEERS-717, this line is too close to the end of the spectral range in our reduction to be robustly detected for our sample. Object CEERS-1236 is initially detected in our broad-line search, but we detect its broad component at S/N$_{b}$=3.47, below our threshold of S/N$_{b}>4$. \cite{harikane23} detected this component at S/N$_{b}$=5.9, but we expect that this difference may be due to our rescaling of the spectral errors by a factor of 1.7 (see \S\ref{sec:extraction}). We also initially detect Object CEERS-1665 as a broad-line candidate, but like CEERS-1236, we detect its broad component at only S/N$_{b}$=3.86: below our S/N$_{b}>4$ cutoff. \cite{harikane23} detected this broad component with S/N$_{b}$=11.6 and FWHM$_{b}$=1794~km~s$^{-1}$, but our fit detects the weak broad component with FWHM$_{b}$=838~km~s$^{-1}$. In a visual inspection of the two fitted spectra, both fits seem to match their respective data well, therefore we attribute this discrepancy to differences in data reduction or continuum modeling (or lack thereof) during the line fitting.

In addition to these analyses of CEERS data, \cite{kocevski24} previously searched the RUBIES data for spectra of LRDs. In their search, they identified 13 \Ha emitting BLAGN: objects RUBIES-EGS-61496, RUBIES-EGS-60935, RUBIES-EGS-49140, RUBIES-EGS-42232, RUBIES-EGS-42046, RUBIES-EGS-37124, RUBIES-EGS-927271, and RUBIES-EGS-926125 in the CEERS field, and objects RUBIES-UDS-50716 , RUBIES-UDS-47509, RUBIES-UDS-44043, RUBIES-UDS-59971, and RUBIES-UDS-63166 in PRIMER-UDS. We detect 11/13 of these objects, and of these 11, we detect eight with S/N$_{b}>4$ (excluding RUBIES-EGS-927271 S/N$_{b}$=1.9, RUBIES-UDS-59971 S/N$_{b}$=3.1, RUBIES-UDS-50716). For the remaining two objects, the 2D spectrum of RUBIES-EGS-61496 is significantly contaminated such that our optimal flux extraction fails and RUBIES-UDS-63166 failed to achieve a S/N$>$5 at its line center pixel. As these objects are likely still BLAGN based on the visual identification of broad components in \cite{kocevski24}, despite failing our selection procedure, we account for BLAGN such as these through our completeness correction (see \S\ref{sec:completeness}). 

Finally, as mentioned above in \S\ref{sec:outflows}, \cite{wang24b} previously identified RUBIES-EGS-49140.
Beyond these previously known objects, we report a new BLAGN in the CEERS dataset: CEERS-11728, and we report 49 new BLAGN in RUBIES (including the confirmation of a PRISM spectrum candidate BLAGN from \citealt{napolitano24}: RUBIES-EGS-28812).

\section{Comparison with Little Red Dots}\label{sec:lrds}

Here, we examine the prevalence and properties of LRDs in our BLAGN sample, and compare our population of LRD BLAGN to our sample of non-LRD BLAGN.

\subsection{UV and Optical Slopes}\label{sec:slopes}
First, we use the above-mentioned photometry (see \S\ref{sec:photometry}) to measure rest-frame UV and optical slopes for each BLAGN. Here, we follow the methodology of \cite{kocevski24} and use a redshift-dependent set of filters to calculate these slopes. Specifically, for objects with $3.25 < z < 4.75$, we use the F814W, F115W, and F150W filters to compute the rest-UV slope, and F200W, F277W, and F356W to compute the rest-optical slope, and for objects with $4.75 < z < 8$ we use F115W, F150W, and F200W, and F277W, F356W, and F444W, respectively. While all of our BLAGN were previously examined photometrically in \cite{kocevski24}, due to the redshift dependence of these measurements, we remeasure the rest-frame UV and optical slopes here using our spectroscopic redshifts instead of the previously-used photometric redshifts. 

\begin{figure}[t]
\centering
\includegraphics[angle=0,width=\columnwidth]{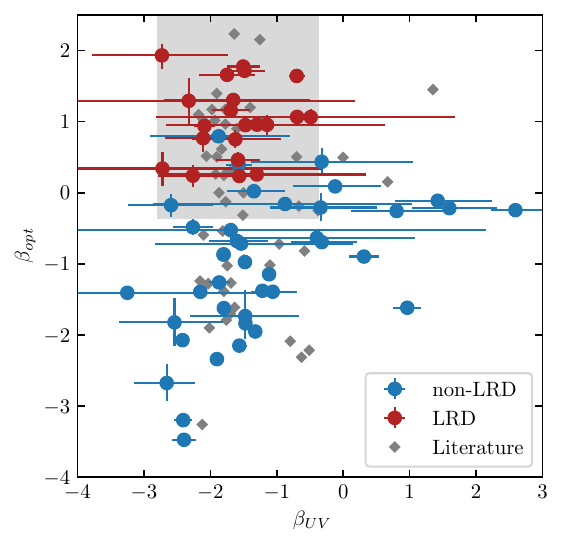}
\caption{Photometrically measured optical slope ($\beta_{opt}$) verses UV slope ($\beta_{UV}$). We show the BLAGN in our sample that are identified as LRDs using the  \citet{kocevski24} criteria as red points, and  BLAGN in our sample that do not satisfy the \citet{kocevski24} selection as blue points. We also show \textit{JWST} identified BLAGN from recent literature \citep{harikane23,matthee24,greene24,juodzbalis24,sun25} as compiled by \cite{hainline25} as grey diamonds. As expected the LRD BLAGN cluster in the shaded region at $-2.8<\beta_{UV}<-0.37$ and $\beta_{opt}>0$, following the \cite{kocevski24} criteria, while the non-LRD BLAGN exhibit a variety of spectral shapes.}
\label{fig:uvopt}
\end{figure}

\begin{figure}[t]
\centering
\includegraphics[angle=0,width=\columnwidth]{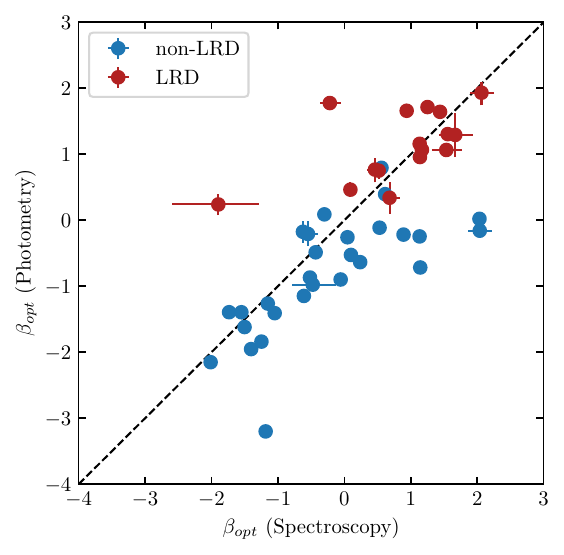}
\caption{Photometrically measured $\beta_{opt}$ versus Spectroscopically measured $\beta_{opt}$. We show the BLAGN in our sample that are identified as LRDs using the \citep{kocevski24} criteria as red points, and BLAGN in our sample that do not satisfy the \citep{kocevski24} selection as blue points. Note that not all of our BLAGN have PRISM data, thus some are excluded from this figure. The black dashed line is the 1:1 line. Objects above this line show a redder $\beta_{opt}$ when measured from photometry instead of spectroscopy likely due to emission line boosting. However, despite the emission line boosting effects, most of the LRDs still exhibit red to moderately red $\beta_{opt}$, implying that they are intrinsically dust-reddened.}
\label{fig:optvopt}
\end{figure}

To perform these fits, we once again use \texttt{emcee} with a simple power law model: $f_{\lambda}= C\lambda^{\beta}$ where $\beta$ is the spectral slope, and $C$ is an arbitrary normalization constant (using 32 walkers and a 5000 step chain, of which we discard the first 3333 steps). In these initial fits, we do not account for any effects of strong emission lines on the broadband photometry (we address this below and in Figure~\ref{fig:optvopt}). Using these slopes and our photometry, we apply the full LRD criteria (including $\beta_{UV}$, $\beta_{OPT}$, and half-light radius cuts, as well as color cuts to reject objects with broadband magnitudes significantly effected by emission lines) from \cite{kocevski24} to determine that our BLAGN sample contains 21 LRDs.  We plot the measured optical slopes ($\beta_{opt}$) verses the measured UV slopes ($\beta_{UV}$) of both the LRD and non-LRD BLAGN populations in Figure~\ref{fig:uvopt}.

We also find three objects that satisfy the LRD $\beta_{UV}$ and $\beta_{opt}$ criteria, but do not meet the full \cite{kocevski24} LRD selection. All three of these objects are classified as non-LRDs due to cuts designed to remove objects with broadband colors that might be significantly effected by strong emission lines (see \citealt{kocevski24}, \S3.1 for details). The remaining object is insufficiently compact in F444W imaging. 
The overlap of our full BLAGN sample with the LRD population is moderate (34\%), indicating clearly that LRDs do not account for all BLAGN at $3.5<z<6.8$. These non-LRD BLAGN objects show diverse UV and optical slopes, likely indicating a variety of reddenings (or lack thereof) of both AGN and nebular emission. For a detailed analysis of reddening of the individual objects in this sample, see \citep{brooks24}.

\cite{kocevski24} express concern that the red $\beta_{opt}$ exhibited by LRDs can be the result of the ``boosting'' of the broadband photometry due to the presence of strong emission lines. As mentioned above, \citet{kocevski24} include criteria in their LRD photometric selections to reduce this effect. Here we test this effect directly using our sample of BLAGN. Using the photometrically flux calibrated PRISM spectra (described in \S\ref{sec:fluxcal}), we fit $\beta_{opt}$ directly to the optical continuum. In this process, we mask the [O\,\textsc{iii}]$\lambda$4363, H$\gamma$, H$\beta$, [O\,\textsc{iii}]$\lambda\lambda$4549,5007, and \Ha emission lines, and fit the same power law model used for the photometric $\beta_{opt}$ fitting to a PRISM spectrum from rest-frame 3645--8000~\AA{} (the same approximate wavelength range---starting at the Balmer break limit---sampled by the photometric fitting scheme introduced in \citealt{kocevski24}). We plot the resulting spectroscopic $\beta_{opt}$ values against the photometrically measured values in Figure~\ref{fig:optvopt}. Due to low S/N in the rest-UV of most of the BLAGN PRISM spectra and the comparative lack of contaminating emission lines, we do not repeat this process to produce spectroscopic $\beta_{UV}$ measurements. 

Figure~\ref{fig:optvopt} shows that only two of the BLAGN LRDs for which we have PRISM spectra show strong signs of emission line boosting. This implies that a significant fraction of the LRDs are indeed intrinsically dust-reddened in the optical beyond the contributions of emission lines to their broadband colors. Interestingly, Figure~\ref{fig:optvopt} also shows that seven objects rejected by \cite{kocevski24} due to possible emission line contamination actually exhibit continua with $\beta_{opt}>0$. As such, photometric definitions and criteria for LRD selection may be both slightly contaminated (as seen by the two out of 16 LRDs that have spectroscopic $\beta_{opt}<0$) and moderately incomplete (as seen by the seven photometric non-LRDs that have spectroscopic $\beta_{opt}>0$). For simplicity, and to best comment on, analyze, and compare to the populations of LRDs identified in photometric searches, we continue to use the photometric \cite{kocevski24} criteria throughout this study, and reserve a dedicated investigation into the properties of spectroscopic LRDs for a future work. 

\subsection{Stacked \Ha Lines}

\begin{figure}[t]
\centering
\includegraphics[angle=0,width=\columnwidth]{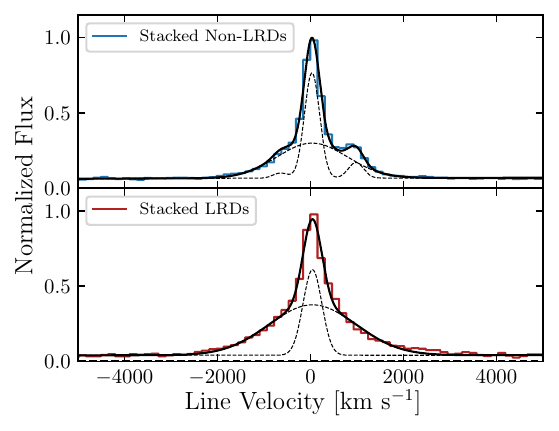}
\caption{Stacked \Ha lines for both the LRDs (red curve) and non-LRDs (blue curve) and fits to each line (broad and narrow components: dashed black curves, combined fits: solid black curves. Both populations show distinct line shapes, where the LRDs show a broader and stronger broad component, while the non-LRDs show a comparatively stronger narrow component and much stronger [N\textsc{ii}] emission. These features suggest that the LRD BLAGN exhibit a distinctly different \Ha line shape and thus relative contributions from the broad-line region and nebular emission than non-LRD BLAGN at the same redshifts.}
\label{fig:hastack}
\end{figure}

We next produce stacks of both the 21 LRD and 42 non-LRD BLAGN \Ha lines to better compare the broad-line properties of the two populations. Here we shift each object's G395M spectrum to the rest frame by dividing both its wavelength and flux values by $(1+z)$, where $z$ is the spectroscopic redshift measured from the peak of \Ha line. We next resample these rest-frame spectra to a common wavelength grid while preserving the observed frame spectral resolution. To best analyze the shape of the \Ha line without biasing our stacks toward the brightest lines, we next re-normalize each object's spectrum to the measured (broad+narrow component) peak of the \Ha line. Finally, we combine the re-normalized spectra with a simple, unweighted, resampled pixel-by-pixel mean (and combine the associated errors in quadrature). We show these stacked \Ha lines in Figure~\ref{fig:hastack}

We next fit these stacked lines using the same two-component \Ha (with [N\textsc{ii}]) model as in the single object fitting. We summarize the results of the fits in Table~\ref{tab:stackfits} and show the fits in Figure~\ref{fig:hastack}. These measurements show that the LRDs tend to have broad components that are both broader ($2200^{35}_{-33}$ versus $1798^{+20}_{-19}$~km~s$^{-1}$) and make up a larger fraction of the total \Ha line emission (73\% vs 59\%) than the broad components of the non-LRDs. Moreover, the LRDs show little to no significant [N\textsc{ii}] emission, while the non-LRDs exhibit a modest log$_{10}$([N\textsc{ii}]$\lambda$6585 / H$\alpha_{narrow}$) of $-0.78^{+0.01}_{-0.01}$. These measurements imply that observations of \Ha in the LRD-selected BLAGN are more significantly dominated by emission from the broad-line region, with lesser contributions from narrow emission than non-LRD BLAGN. 

We repeat this stacking process (this time using inverse variance weighting and no normalization) for the [O\,\textsc{iii}$\lambda\lambda5007,4959$] doublet to retest both the LRD and non-LRD populations for the presence of outflows. We apply the same fitting procedures as described in \S\ref{sec:outflows}, and find that neither stack shows evidence for outflows. Specifically, we measure a $\Delta$BIC=4.9 in favor of a outflow component model for the [O\textsc{iii}] doublet in the LRD stack with a outflow S/N=0.34. The non-LRD stack exhibits a $\Delta$BIC=1.6 in favor of an outflow component, and the component is also insufficiently detected with a S/N=0.42. As such, we do not find significant evidence for outflows in our LRD or non-LRD samples.

\begin{deluxetable}{ccc}
\tablecaption{H$\alpha$ Stack Properties \label{tab:stackfits}}
\tablehead{Quantity & non-LRD Stack & LRD Stack}
 \startdata
Number of Objects & 41 & 21 \\
Narrow Flux Fraction & $0.395^{+0.005}_{-0.005}$ & $0.269^{+0.007}_{-0.007}$ \\
Broad Flux Fraction & $0.605^{+0.005}_{-0.005}$ & $0.731^{+0.007}_{-0.007}$ \\
Broad FWHM [km s$^{-1}$] & $1768^{+18}_{-17}$ & $2188^{+33}_{-33}$ \\
log$_{10}$([N\textsc{ii}]$\lambda$6585 / H$\alpha_{n}$) & $-0.82^{+0.01}_{-0.01}$ & $-3.02^{+0.41}_{-0.61}$
\enddata
\tablecomments{The reported FWHM of the broad component is corrected for instrumental broadening. [N\textsc{ii}] is undetected in the LRD stack, so we give the 1$\sigma$ upper limit for log$_{10}$([N\textsc{ii}]$\lambda$6585 / H$\alpha_{n}$). H$\alpha_{n}$ refers to the narrow component of the \Ha line.}
\end{deluxetable}

\section{The Black Hole Mass Function}\label{sec:BH mass function}

We next use our BLAGN sample to compute the BH mass function. We compute black hole masses for each object from the broad component luminosity ($L_{H\alpha,b}$) and broad component FWHM ($\textrm{FWHM}_{H\alpha,b}$) of the \Ha line using the empirical relations from \cite{reines13}. Specifically, we use their equation (5), duplicated below, where $L_{H\alpha,b}$ and $\textrm{FWHM}_{H\alpha,b}$ are the luminosity and FWHM of the broad component of \Ha, respectively. 

\begin{equation}
\label{eq:mbh_ha}
\begin{aligned}
    M_{\textrm{BH}} = 10^{6.57} & \times \left(\frac{L_{H\alpha,b}}{10^{42}\textrm{~ergs s}^{-1}}\right)^{0.47} \\
    &\times \left(\frac{\textrm{FWHM}_{H\alpha,b}}{10^{3}\textrm{~km s}^{-1}}\right)^{2.06} M_{\odot}
\end{aligned}
\end{equation}

Unlike its predecessor from \cite{greene05} which used the luminosity of both the broad and narrow components of the \Ha line, this equation relies only on the properties of the broad component of the \Ha line, and is thus ideal for computing BH masses for our BLAGN sample. We compute the broad \Ha luminosities for our sample using the sample's spectroscopic redshifts and our above-stated flat cosmology. To best propagate the uncertainties in our \Ha line fitting, we compute BH masses for our samples using the full flux and instrumental-broadening corrected FWHM posteriors from our MCMC fitting, and thus obtain posterior distributions of the BH masses that fully account for any covariance between the luminosity and FWHM of the broad \Ha lines. We caution that the \cite{reines13} formula is calibrated at $z<0.06$, and may not be accurate in the high redshift universe. However, as no such calibration has yet been done at high redshift, we use the \cite{reines13} calibration regardless. In light of this, the uncertainties inherent in the \cite{greene05,reines13} relations dominate over the uncertainties derived from our measurements and are approximately 0.5~dex \citep{reines15}. When applied at higher redshift, these uncertainties may increase further (e.g. deviations of $\sim 0.7$~dex are measured in \citealt{abuter24}). 

\begin{figure}[t]
\centering
\includegraphics[angle=0,width=\columnwidth]{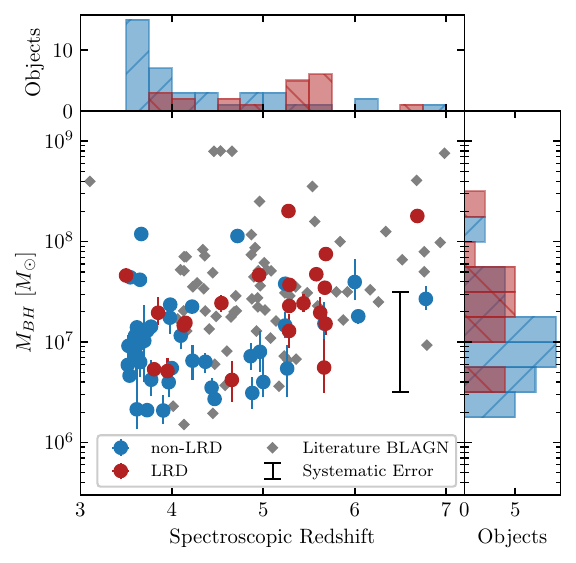}
\caption{Black Hole mass $(M_{\textrm{BH}})$ plotted as a function of spectroscopic redshift for the BLAGN sample. We show objects that are identified as LRDs in \citet{kocevski24} as red points, and objects that do not satisfy the \citet{kocevski24} selection as blue points. The error bars here show the 16th and 84th percentiles of the $(M_{\textrm{BH}})$ posterior distributions, and do not take into account the systematic uncertainties from the \cite{reines13} prescription ($\sim0.5$~dex, shown as the black point and error bar). We also plot histograms of the distributions of LRD and non-LRD BLAGN as functions of spectroscopic redshift and $(M_{\textrm{BH}})$ along the edges of the plot. The BLAGN show no strong correlation between $(M_{\textrm{BH}})$ and redshift, implying few to no redshift-dependent biases in the sample selection. However, the population of LRD BLAGN exhibit significantly higher redshifts and $(M_{\textrm{BH}})$ than the non-LRD BLAGN. We plot the combined \textit{JWST} H$\alpha$ identified BLAGN from \cite{carnall23,harikane23,Maiolino_2023,greene24,juodzbalis24,Killi23,kocevski24,Lin_2024,ubler24}, and \cite{wang24a} as grey diamonds.}
\label{fig:bhmz}
\end{figure}

We further verify our black hole masses by re-deriving them using the $L_{5100}$ prescription from \cite{greene05}. Here, we combine their equations (3) and (5) to derive an expression (our equation~\ref{eq:mbh_5100}) for black hole mass as a function of continuum luminosity at rest-frame 5100~\AA{} ($L_{5100}$) and the FWHM of the broad component of \Ha ($FWHM_{\textrm{\Ha},b}$). We measure $L_{5100}$ directly from the PRISM spectra (when available) and use our fitted values of FWHM$_{\textrm{\Ha},b}$ in this expression. We find that the $L_{5100}$ derived black hole masses are well-matched to the $L_{\textrm{\Ha}}$ black hole masses with a median offset $\log_{10}\left(M_{BH,L\textrm{\Ha}}/M_{BH,L5100}\right)=$0.13 and an absolute median deviation of 0.28. As this absolute median deviation is still significantly less than the assumed $\sim0.5$~dex uncertainty inherent in these $M_{\textrm{BH}}$ prescriptions, we consider this to be good agreement. With this confirmation, and our lack of PRISM data for some objects, we use our $L_{\textrm{\Ha}}$ black hole masses in the remainder of this work. 

\begin{equation}
\label{eq:mbh_5100}
\begin{aligned}
    M_{\textrm{BH}} = & (4.71\pm 0.17) \\
    & \times 10^{6}\left(\frac{L_{5100}}{10^{44}\textrm{~ergs s}^{-1}}\right)^{0.64\pm0.02} \\
    &\times \left(\frac{\textrm{FWHM}_{H\alpha,b}}{10^{3}\textrm{~km s}^{-1}}\right)^{2.06\pm0.06} M_{\odot}
\end{aligned}
\end{equation}

We next estimate the effects of dust reddening on our computed BH masses using the methodology from \citet{kocevski24} (see their \S3.3 for full details). 
Briefly, we use a custom $\chi^2$ fitter to fit composite AGN and star-formation models to the photometric data for each source. From these fits, we derive an approximate reddening $(A_V)$ for each AGN. We caution that these fits may be significantly skewed by the presence of strong emission lines and assume that the AGN component dominates the reddened rest-frame optical continuum emission. As such, we use these fits as simple estimates of $A_V$. We next assume a \cite{calzetti00} dust law (with $R_V=4.05$), correct our \Ha fluxes for this reddening, and recompute $M_{\textrm{BH}}$ for each object. Based on these $A_V$ estimates, we find that the median increase in $M_{\textrm{BH}}$ for the BLAGN sample is 0.19~dex, with a maximum increase of 0.55~dex for the most reddened object in the dataset. This effect is most prominent for the more massive objects in the dataset, where objects with $M_{\textrm{BH}}<10^7 M_{\odot}$ have a median correction of 0.05~dex, and objects with $M_{\textrm{BH}}<10^7 M_{\odot}$ have a median correction of 0.31~dex. While these corrections offer an estimate of the un-attenuated \Ha fluxes, due to the uncertainty in our SED modeling and fitting, and the complexity of the SEDs of LRDs and BLAGN as a whole, we again continue to use our uncorrected BH masses and \Ha fluxes in the remainder of this work with the caveat that the BH masses of our BLAGN population may be underestimated by $\sim0.05^{+0.20}_{-0.18}$~dex at $M_{\textrm{BH}}<10^7 M_{\odot}$ and by $\sim0.31^{+0.11}_{-0.23}$~dex at $M_{\textrm{BH}}>10^7 M_{\odot}$. For a detailed analysis of reddening of the individual objects in this sample, see \cite{brooks24}. In brief, \cite{brooks24} examine this sample, as well as \cite{harikane23, kocevski23}, and \cite{maiolino23b}, and find that the broad-line component of H$\beta$ is undetected ($<3\sigma$) in over 86\% of \textit{JWST} identified BLAGN. Therefore, it is very difficult to constrain the reddenings of the broad-line regions of these objects using Balmer lines \citep[see][their Figure 4]{brooks24}. Nonetheless, our analysis above indicates that the effect of reddening on our computed BH masses ($\sim$+0.19~dex) is small compared to the $\sim\pm$0.5~dex uncertainty from the \cite{reines13} prescription.

We plot the computed BH masses as a function of object redshift in Figure~\ref{fig:bhmz} (with the 0.5~dex systematic uncertainty depicted as a black error bar) and list the computed masses in Table~\ref{tab:sample}. Notably, the BH masses estimated here are in broad agreement with those found in other collections of \textit{JWST}-detected BLAGN at $3.5<z<7$ \citep[e.g.,][]{carnall23,harikane23,Maiolino_2023,greene24,juodzbalis24,Killi23,kocevski24,Lin_2024,ubler24,wang24a} ranging from $10^{6}-10^{9} M_{\odot}$. We also note that while our sample of LRDs span roughly the same range of BH masses as the non-LRD sample, as a population, the LRDs exhibit greater BH masses $\left(\log_{10}\left(M_{\textrm{BH}}/M_{\odot}\right)=7.4^{+0.3}_{-0.6}\right)$ than the non-LRDs $\left(\log_{10}\left(M_{\textrm{BH}}/M_{\odot}\right)=7.0^{+0.5}_{-0.4}\right)$. However, as there are only 21 LRDs in our sample, our sample is still too small for a robust statistical comparison of these mass distributions.

To derive our selection volume, we require a measurement of our survey area. For this, we use the full areas of the four quadrants of the NIRSpec MSA for each pointing to determine the field areas (shown as black rectangles in Figure~\ref{fig:fields}), and after accounting for overlap between NIRSpec pointings, we determine a total probed area of 181.98~arcmin$^2$ for the 24 NIRSpec MSA pointings used in this study (six from CEERS, six from RUBIES-EGS, and 12 from RUBIES-UDS). We choose a redshift range of $(3.5<z<6)$ for our BH mass function that encompasses the main distribution of BLAGN in our sample (this excludes the two highest redshift sources at $z=6.685$ and $z=6.778$). The co-moving volume over this redshift range and area is 1.60 $\times$ 10$^{6}$ Mpc$^3$. 

\subsection{Sample Completeness}\label{sec:completeness}

To calculate the black hole mass function, we must understand the overall completeness of our BLAGN selection.  To do this, we compute two different and independent sources of incompleteness: ``line detection'' and ``observational''. We define ``line detection'' incompleteness to be incompleteness due to observed BLAGN that are not detected by our broad-line finding algorithms, or are discarded due to our broad-line selection cuts. Hence, despite these sources being present in our observed sample, we did not detect their broad-lines. 

\begin{figure*}[t]
\centering
\includegraphics[angle=0,width=\textwidth]{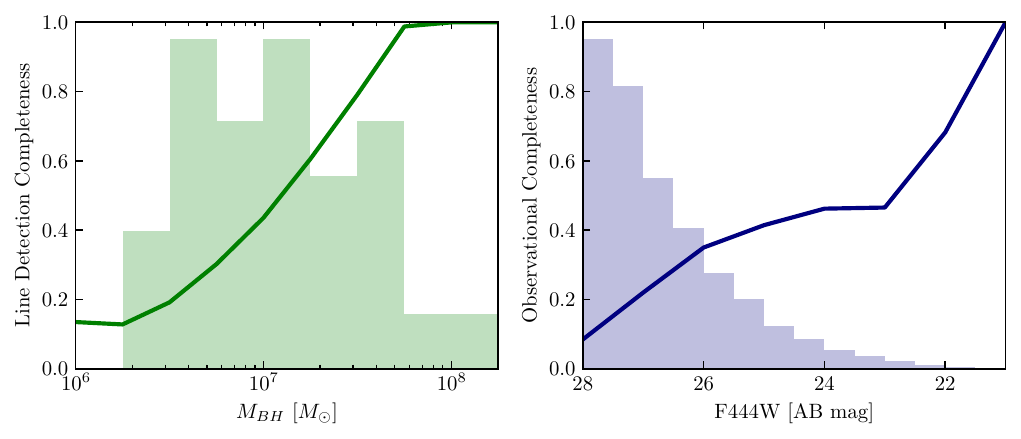}
\caption{\textit{Left:} Line Detection Completeness plotted as a function of Black Hole Mass $(M_{\textrm{BH}})$ (green curve). We plot the distribution of BH masses in our $3.5<z<6$ BLAGN sample as the light green histogram (arbitrary normalization). The completeness curve increases with $\log_{10}\left(M_{\textrm{BH}}\right)$ due primarily to the S/N of the broad component generally increasing with $M_{\textrm{BH}}$. \textit{Right:} Observational Completeness plotted as a function of F444W magnitude (dark blue curve). We plot the distribution of F444W magnitudes of the sources (both BLAGN and other objects) observed by CEERS and RUBIES as the faded blue histogram (arbitrary normalization). Observational completeness generally increases with F444W brightness for two intertwined reasons: bright objects are both intrinsically rarer and simultaneously easier to observe, thus a greater fraction of these objects were targeted in both CEERS and RUBIES.}
\label{fig:comp}
\end{figure*}

To quantify the prevalence of these missed sources, we use a simulation using the \texttt{Pandeia} \textit{JWST} ETC engine \citep{pontoppidan16} to produce simulated \textit{JWST} NIRSpec MSA G395M spectra featuring modeled broad \Ha lines. These modeled \Ha lines use a seven parameter model consisting of: narrow \Ha flux, broad \Ha flux, narrow \Ha FWHM, broad \Ha FWHM, [N\textsc{ii}]$\lambda6585$/\Ha ratio, \Ha equivalent width (EW, computed with the sum of the broad and narrow components, and used to set the continuum flux level), and redshift. Critically, this combination of parameters also inherently includes a derived BH mass for each synthetic source. We vary these parameters on a large seven-dimensional grid and produce a sample of broad \Ha lines that span the physical parameters of our BLAGN sample. We sample narrow \Ha FWHM from 1--750~km~s\per, broad \Ha FWHM from 500--3000~km~s\per, [N\textsc{ii}]$\lambda6585$/\Ha ratio from 0--0.5, and redshift from 3.5--6.0 as flat distributions in linear space. We sample narrow and broad \Ha flux from $10^{-19}-10^{-16}$~ergs~s$^{-1}$~cm$^{-2}$ and \Ha rest-frame equivalent width from 100--1000~\AA{} as flat distributions in logarithmic space. 
While this grid may not be representative of the true underlying distribution of BLAGN, as we do not have any strong priors on the \Ha line properties of BLAGN at these redshifts and line fluxes outside of our own detected sample, we use the above simple flat priors grid as our best option, and simulate $\sim150,000$ objects.

We adopt the RUBIES NIRSpec observation strategy and parameters and use the modeled \Ha spectra as input spectra for \texttt{Pandeia} to produce simulated observed spectra. We then analyze this sample of synthetic spectra using the same broad-line detection/fitting algorithms and parameters as described in \S\ref{sec:linedetection}. 

We derive our detection completeness by dividing our synthetic sample into bins of width $\Delta \log_{10}\left(M_{\textrm{BH}}/M_{\odot}\right)=0.25$, and determining the fraction of the synthetic objects in each bin that are successfully detected and determined to be BLAGN based on their \Ha line by our selection algorithms. We show the overall results of these tests in Figure~\ref{fig:comp} (left panel). 

We find that the dominant variables in determining an object's detection are the flux of the narrow and broad \Ha components, while the redshift of the source, the [N\,\textsc{ii}]/\Ha ratio and the \Ha EW have minimal to no effect. This is fully expected, as in order to be initially detected, an object must have a peak pixel S/N$>5$, which is dependent on both the broad and narrow component fluxes. Then, the object must ultimately fit with a broad component S/N$>4$, which depends directly on the broad component flux. These effects are also coupled to the FWHM of the broad and narrow components, such that broader lines will be harder to detect, as the flux is spread over larger areas of the spectrum and thus incur a greater amount of read-noise and a lower S/N on the line peak. This effect explains why the completeness trends steadily downwards with decreasing $M_{\textrm{BH}}$ instead of dropping suddenly from near 100\% to near 0\% at a cutoff $M_{\textrm{BH}}$ corresponding to a strict S/N cut. 

We also correct our sample for ``observational'' incompleteness. Here we define ``observational'' incompleteness as incompleteness due to candidate sources that fall within the MSA pointings areas but were not assigned open shutters. To characterize this sample, we first select all objects in the photometric catalogs that fall within the observed MSA pointings. We next use the photometric properties of our confirmed BLAGN sample to select objects in the photometric catalog that could be candidate BLAGN. We require that catalog objects are brighter than the faintest BLAGN in both F356W and F444W, and have colors F277W--F444W and F150W--F444W that are redder than the bluest objects in the BLAGN sample. Finally, we require that the best-fit photometric redshifts of the catalog objects lie within our selected redshift range: $3.5<z<6$. These cuts produce a sample of candidate objects with photometric properties similar to our observed BLAGN sample.

To use this candidate sample to constrain our observational completeness, we sort the candidate and confirmed BLAGN samples into bins of F444W magnitude with a bin width of 1 magnitude. As we cannot derive BH masses for the photometric candidate objects, we instead use F444W to compare to the BLAGN sample. In each bin, we determine the completeness as the fraction of candidate objects (both BLAGN and other objects) that were assigned slits and observed in G395M. We plot our observational completeness curve in Figure~\ref{fig:comp} (right panel).

To apply these completeness measures to our sample---for a given BLAGN---we interpolate the line detection and observational completeness curves to an object's $M_{\textrm{BH}}$ and F444W magnitude respectively. When multiplied together, these values produce the overall completeness for an object. In constructing the BH mass function (see below), we use one over this overall completeness as a weight for the given object. For example: a BLAGN with a $M_{\textrm{BH}}=10^7~M_{\odot}$ and a F444W magnitude of 26 would have a line detection completeness of $\approx0.5$ and an observational completeness of $\approx0.4$. This implies that the \Ha lines of BLAGN with similar $M_{\textrm{BH}}$ to this object are detected $\approx$50\% of the time in the observed sample, and $\approx$40\% of objects of a similar brightness have been observed in the surveyed area. As such, only $\approx$20\% of objects like this one have been both observed and detected as BLAGN. Therefore, we up-weight this object by a factor of $\approx$5 to account for the similar objects that are likely missed in the survey volume.

\begin{figure*}[t]
\centering
\includegraphics[angle=0,width=\textwidth]{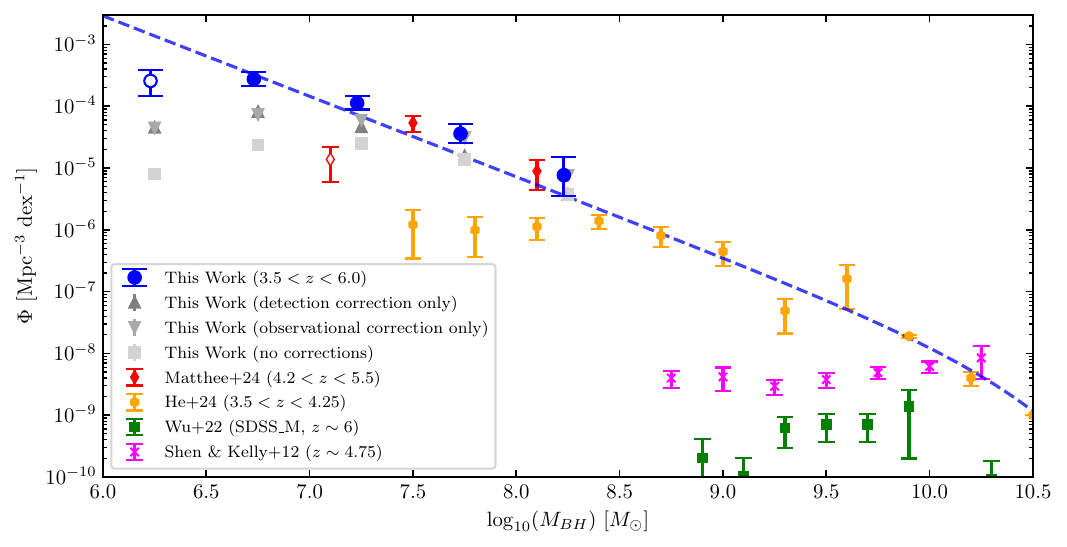}
\caption{The BLAGN BH mass function for $3.5<z<6$. Blue points indicate the fully completeness-corrected BH mass function from this work, and upward/downward pointing grey triangles indicate our BH mass function with only the line detection/observational correction applied. We mark our lightest mass point with a hollow blue point to indicate its low line detection completeness of $<20\%$. Grey squares indicate our uncorrected BH mass function. All of our error bars are Poissonian (following the \citealt{gehrels86} prescription). Our partially corrected or uncorrected BH mass functions share the same error bars (in logarithmic space) as our fully corrected BH mass function and are omitted to reduce visual clutter. Red diamonds show the BH mass function derived from NIRCam WFSS data in \cite{matthee24}. Note that \cite{matthee24} acknowledge that their lightest mass point (marked with a hollow red diamond) likely suffers from incompleteness and does not indicate a turnover of the BH mass function at $M_{\textrm{BH}}\sim10^7~M_{\odot}$. Yellow hexagons show the BH mass function at $3.5<z<4.25$ derived from ground-based SDSS+HSC data in \cite{he24}. Green squares show the BH mass function at $z\sim6$ from \cite{wu22}, and pink x's show the BH mass function from \cite{shen12}, both derived from SDSS data. We show a Schecter fit to our data from $10^{6.5}<M_{\textrm{BH}}<10^{8.5}$ and the \cite{he24} data at $M_{\textrm{BH}}>10^{8.5} M_{\odot}$ as the dashed blue curve. Our corrected points show excellent agreement with the \cite{matthee24} points indicating that our observational completeness correction is performing well, and our overall BH mass function seems to extend the \cite{he24} BH mass function to lower masses.}
\label{fig:BHmassfunction}
\end{figure*}

\begin{deluxetable*}{cccccc}[htb]
\tablecaption{Black Hole Mass Function \label{tab:BHmassfunction}}
\tablehead{$M_{\textrm{BH}}$ & $\Phi$ & N & \multicolumn{3}{c}{Completeness} \\ $\left[\log_{10}\left(\frac{M_{\textrm{BH}}}{M_{\odot}}\right)\right]$ & [$10^{-6}$ Mpc$^{-3}$dex$^{-1}$] & & Total & Obs. & Det.}
\startdata
6.25 & $258^{+125}_{-113}$ & 6.45 & 0.03 & 0.18 & 0.18 \\
6.75 & $276^{+83.7}_{-59.8}$ & 18.74 & 0.09 & 0.32 & 0.29 \\
7.25 & $113^{+32.5}_{-24.3}$ & 19.84 & 0.22 & 0.42 & 0.53 \\
7.75 & $36.1^{+14.8}_{-10.5}$ & 10.93 & 0.38 & 0.44 & 0.87 \\
8.25 & $7.67^{+7.51}_{-4.17}$ & 2.99 & 0.49 & 0.49 & 1.00 \\
\enddata
\tablecomments{Columns: $M_{\textrm{BH}}$ is the log-center mass of the BH mass function bin, $\Phi$ is the number density (per dex of $M_{\textrm{BH}}$) of BLAGN in each BH mass bin, N is the number of BLAGN in each bin (the non-integer values are due to our distributing each BLAGN fractionally into the $M_{\textrm{BH}}$ bins according to the posterior distribution of its fitted $M_{\textrm{BH}}$), Total Completeness is the average of the products of observational and line detection completeness for each object in the mass bin, Obs.\ Completeness is the fraction of objects in a mass bin that were observed in the NIRSpec MSA observations (see \S\ref{sec:completeness}), and Det.\ Completeness is the fraction of objects in a mass bin whose \Ha broad-lines are detected in our completeness simulations (see \S\ref{sec:completeness}).}
\end{deluxetable*}

\subsection{The Black Hole Mass Function}\label{sec:subBH mass function}

Using our completeness corrected BLAGN sample and the computed comoving volume, we construct the BH mass function in Figure~\ref{fig:BHmassfunction} and show the associated data in Table~\ref{tab:BHmassfunction}. We define bins of $\log_{10}(M_{\textrm{BH}})$ from $\log_{10}(M_{\textrm{BH}})=6-9$ $M_{\odot}$ with binwidths of 0.5 dex. We use the saved posterior distribution of $M_{\textrm{BH}}$ of the sample to account for the uncertainties on each object's $M_{\textrm{BH}}$ by distributing each BLAGN fractionally into the $M_{\textrm{BH}}$ bins according to the posterior distribution. Finally, we construct the BH mass function by dividing the weighted object count in each bin by the bin width and the co-moving volume. We assume Poissonian errors for each bin following \cite{gehrels86}. 

We plot our final $3.5<z<6$ BH mass function in Figure~\ref{fig:BHmassfunction} (blue points) alongside the \textit{JWST} based BH mass function from \cite{matthee24} (red diamonds). The \cite{matthee24} BH mass function ($4.2<z<5.5$) is constructed from a sample of BLAGN identified in NIRCam Wide Field Slitless (WFSS) data from the EIGER and FRESCO programs \citep{kashino23,oesch23}. Because these data are from slitless spectroscopy, no observational completeness correction is required. As such, our BH mass function's strong agreement with the \cite{matthee24} data indicates that our observational completeness correction is performing well. We do note that the \cite{matthee24} data are also not corrected for line detection effects, however, our line detection completeness is $>80\%$ at $M_{\textrm{BH}}>10^{7.5} M_{\odot}$ so a direct comparison is still appropriate. Figure~\ref{fig:BHmassfunction} also demonstrates the superior depth of the CEERS and RUBIES NIRSpec MSA data over the WFSS, as our BH mass function probes BHs nearly an order of magnitude less massive than the lightest WFSS detected BLAGN. While previous ground-based studies \citep[e.g.]{shen12,wu22} were limited to $M_{\textrm{BH}}\gtrsim 10^9 M_{\odot}$, it is clear that \textit{JWST} has revealed the BH mass function at masses that ground-based data \citep[e.g.][$3.5<z<4.25$]{he24} have only just begun to probe. Our BH mass function seems to follow an extrapolation of the \cite{he24} mass function to lower masses with a power-law like shape. We fit a Schecter function \citep{schechter76} to our data and the \cite{he24} data at $M_{\textrm{BH}}>10^{8.25} M_{\odot}$ (shown as a dashed blue curve in Figure~\ref{fig:BHmassfunction}) and derive the following parameters: $\Phi^*=5.32^{+4.32}_{-2.76}\times 10^{-9}$~Mpc$^{-3}$~dex\per, $M^*=2.55^{1.39}_{-0.72}\times 10^{10} M_{\odot}$, and $\alpha=-1.30^{+0.04}_{-0.04}$, affirming this power-law-like shape out to nearly the heaviest end of the \cite{he24} data. Our BH mass function may show signs of flattening out at the lightest masses $M_{\textrm{BH}}<10^7 M_{\odot}$ when compared to the light-end power law from of the Schecter fit, but the significance of this flattening is uncertain given our reduced completeness at these low masses. This flattening could be better examined with deeper observations to improve our sensitivity to BLAGN with less massive BHs. 

\subsection{Comparison with Empirical and Theoretical Models}\label{sec:models}

\begin{figure*}[t]
\centering
\includegraphics[width=\textwidth]{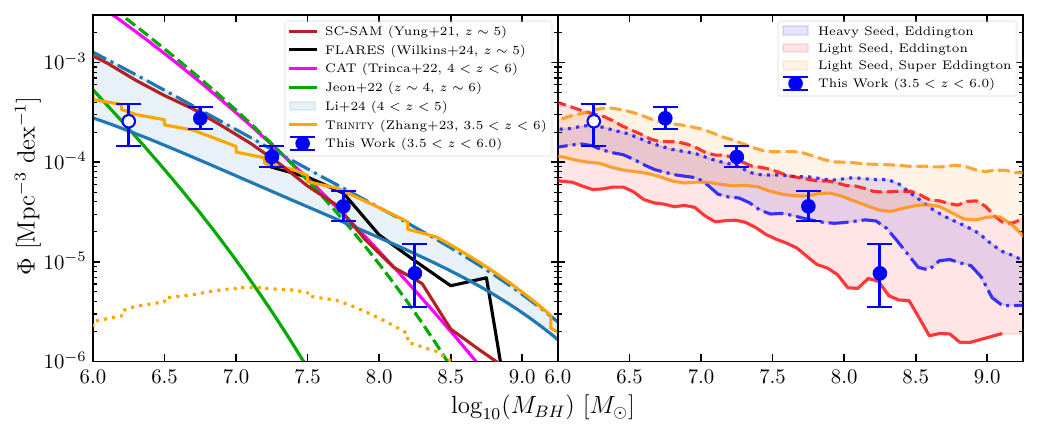}
\caption{Theoretical and empirical modeling of $z\sim 4 - 6$ BH mass function. In both panels, we reproduce our newly derived observed mass function from Fig.~\ref{fig:BHmassfunction}. \textit{Left panel:} Comparison with select theoretical predictions from the literature at $z\sim4-6$. We show the mass functions averaged over $z \sim 4-5$ from \cite{Li2024}, where the solid blue curve is their model seeded with $N_{\rm seed}=10^{-3}$~Mpc$^{-3}$, and the dot-dashed blue curve seeded with $N_{\rm seed}=10^{-2}$~Mpc$^{-3}$. The analytical model of \citet{Jeon2022} is shown for $z=4$ and $z=6$ as the solid and dashed green curves. We further show the Eddington limited (``reference'') BH mass function of \cite{Trinca2022} at $z=4-6$ as the solid pink curve, the BH mass function from the First Light and Reionization Epoch Simulations (FLARES, black curve; \citealt{wilkins24}), and from the Santa Cruz semi-analytic model (SC-SAM, dark red curve; \citealt{yung21}). Finally, we show BH mass functions for all galaxies (solid orange curve) and for only BLAGN (dotted orange curve) from the T\textsc{rinity} semi-analytic model \citep{zhang23}. We note that the FLARES model exhibits very strong agreement with our observed BH mass function at the heavy end, and the SC-SAM shows strong agreement over our full range of BH masses. \textit{Right panel:} Results from our idealized merger-tree modeling at $z\sim 5$, as described in the Appendix. The shaded regions (orange, brown, and red) reflect the variation (see Appendix) in light seed mass (100 to $10^4 M_\odot$; solid and dashed curves) and heavy-seed (occupation) fraction (1 to 10\%, dot-dashed and dotted curves). According to this modeling, it is evident that details of the seeding mechanism do not make a difference by the time considered here ($z\sim 5-6$).
}
\label{fig:BHmassfunction_theory}
\end{figure*}

We next reproduce select theoretical predictions at $z>4$ from the literature in Figure~\ref{fig:BHmassfunction_theory}, left panel, (for a review of theoretical predictions at $z<4$, see \citealt{habouzit21}). These models from the literature employ a variety of approaches to model the formation of initial seed BHs, their subsequent growth, and the impact of BH feedback on the BH growth itself as well as on the galaxy.

Here, we briefly summarize the key aspects of this modeling that are most relevant for this work, and refer the reader to the original studies for details. First, we show results from two semi-analytic models of the co-evolution of galaxies and SMBH, which are based within dark matter merger histories. The Santa Cruz semi-analytic model \citep[SC-SAM;][]{somerville08,hirschmann2012,yung21} populates top level halos with seeds of mass $10^4 M_\odot$, where top level halos are 0.01 times the mass of the root halo or $10^{10} M_\odot$, whichever is smaller. Rapid black hole growth is triggered by mergers and disk instabilities, where the growth rate is capped at the Eddington rate. Following a triggering event, BHs grow at the Eddington rate until they reach a critical mass, above which BH feedback halts further growth. BHs can also grow by merging. The Cosmic Archaeology Tool (CAT) semi-analytic model \citep{Trinca2022} explored different combinations of BH seeding and accretion scenarios, including both light (Pop III) and heavy (direct collapse; $M_{\rm seed} \sim 10^5\,M_{\odot}$) seeds, and both Eddington-limited accretion and a model in which brief periods of super-Eddington accretion occur following galaxy mergers. 
We also show results from two semi-empirical/analytic models. \citet{Li2024} seed BHs at $z=20$ and apply a parameterized stochastic accretion model. They calibrated the parameters controlling seeding and accretion to reproduce the observed quasar luminosity function at lower redshifts. We further consider the distribution of BH masses derived from a Press-Schechter halo mass function, employing a conversion efficiency between halo and BH masses, which in turn assumes that the BH feedback energy is balanced by the gravitational potential energy of the host system \citep{Jeon2022}.
Additionally, we show the BH mass function from the FLARES cosmological hydrodynamical simulation \citep{wilkins24}. BH seeds with a mass of $10^5$ h$^{-1}\, M_{\odot}$ are inserted into halos with masses greater than $10^{10} M_{\odot}$, and accretion is modeled with an Eddington-limited Bondi-Hoyle approach. BHs can also grow by merging. Thermal energy associated with black hole accretion is injected stochastically into particles surrounding the BH. 
Finally, we show the BH mass function from the empirical T\textsc{rinity} model \citep{zhang23}. T\textsc{rinity} parametrizes the dark matter halo—galaxy--SMBH connection from z=0-10 as a series of redshift-dependent scaling relations, and finds the best fitting parameters that simultaneously match a large compilation of galaxy data from $z=0-13$ and SMBH data from z=0-6.5. From the best-fitting T\textsc{rinity} model, we extract the SMBH mass—luminosity distributions to forward model the BLAGN mass functions.

All of the above mentioned models and simulations predict a rather featureless, power law BH mass function over the observed range of BH masses. Even more strikingly, the three most detailed models (the SC-SAM, CAT, and FLARES) reproduce the observed BH mass function shape and normalization quite well over this mass range, and the other models are also in the right general ballpark. This is significant because many models have difficulty growing large enough populations of the more massive black holes at $z\sim 6-7$ (with masses of a few billion $M_\odot$) from physically motivated seed masses with Eddington-limited accretion. This comparison shows that, although the newly discovered population of lower mass BHs uncovered by JWST is often described as surprising or unexpected, it is actually consistent with theoretical expectations from several models \citep[see also:][]{ricate18}. However, this consistency comes with the caveat that all of the above BH mass function models show the total BH mass function consisting of all galaxies, not just those hosting BLAGN. As a comparison, we show a modified BH mass function in Figure~\ref{fig:BHmassfunction_theory} (orange dotted curve) from T\textsc{rinity} that only shows the contribution of BLAGN to the BH mass function. This BLAGN-only BH mass function shows a 1-2~dex separation from our observed BLAGN BH mass function, suggesting that---especially at the low-mass end---in at least this particular model, the BLAGN mass function may underestimate either the active fraction of SMBHs or the total mass function when compared to \textit{JWST}-based observations.

\subsection{Exploring Black Hole Seeding with a Toy Model}

Our derived BH mass function is largely featureless, with a power-law-like functional form. This suggests that the galaxy formation process has wiped out any memory of the specific seeding of the central supermassive black hole (SMBH) already by $z\sim 5$. This further allows us to place constraints on the redshifts that have to be reached to discriminate between the different seeding models (in this regard, see also the methodology presented in \citealt{Pacucci_Loeb_2022}). In the following, we will briefly consider this timing constraint, with an idealized, back-of-the-envelope argument, to be followed up with a dedicated investigation in future work.

What is the time needed to transform a narrowly peaked input mass function, centered around a given BH seed mass $m_0$ (light or heavy, see above), into a featureless power law (see \citealt{Fragione_2023} for a similar, Bayesian analysis of quasars at $z > 6$)? Accretion will change the BH masses, resulting in a BH number density, $n(m,t)$, whose evolution is governed by a continuity equation \citep[e.g.,][]{Oey2011}:
\begin{equation}
 \frac{\partial n}{\partial t}+\frac{\partial}{\partial m}\left(\dot{m}n\right)=0\mbox{\ ,}   
\end{equation}
where $\dot{m}$ is the BH accretion rate. For simplicity, we here assume that this rate is centered around a mean value of $m_0/t_{\rm Sal}$, where the Salpeter timescale, $t_{\rm Sal}\simeq 5\times10^7$\, yr, is the $e$-folding time for Eddington-limited growth \citep[e.g.,][]{Haiman2001}. We further assume that the dispersion in the accretion rate, due to variations in the SMBH environment, is of the same order as the mean, $\sigma_{\rm acc}\simeq m_0/t_{\rm Sal}$. The continuity equation can then be re-written to give an estimate for the time needed to establish a given spread in BH masses, say $\Delta m\simeq 10m_0$, owing to the spread in accretion rate: $\Delta t\sim \Delta m/\dot{m}\sim (10 m_0/m_0)\times t_{\rm Sal}\sim 10 t_{\rm Sal}\sim 5\times 10^8$\, yr. Given that we observe the featureless BH mass function at $z\sim 5$, we can estimate that any signature for a specific seeding mechanism would only show up at $z\gtrsim 10$. This, then, is the target epoch for future, even deeper observations with \textit{JWST}, aimed at elucidating the nature of the first (seed) black holes.

While this loss of memory can be demonstrated from the literature models, the variety of different feedback, accretion, and other prescriptions in these models makes direct comparison of different seeding schemes difficult. Therefore, we further illustrate this loss of memory of the initial seeding process with a simple idealized model, where we insert seed black holes into a halo merger tree and trace their growth for a few select scenarios.  The full model is described in the Appendix, but in brief, we insert both light ($10^{2-4} M_{\odot}$) and heavy ($10^5 M_{\odot}$) seeds into the merger trees from the THESAN-DARK-2 simulation output \citep{Geraldi2022,Kannan2022,Smith2022}.  These black holes grow with their halos following prescriptions from the literature, and we derive the black hole mass function at the observed epoch for light seeds growing at both Eddington and super-Eddington rates, and heavy seeds growing at the Eddington limit. We show outputs for these models in Figure~\ref{fig:BHmassfunction_theory} (right panel). The different scenarios exhibit significant overlap, again indicating that the BH mass function at $z\sim 5-6$ is not sensitive to the specifics of the seeding process at earlier times.

In addition to these toy models, our selection of models from the literature, all assuming various feedback-regulated prescriptions (see above), further illustrate the loss of any signature of the initial seeding process. Specifically, all of these models roughly reproduce the near power-law behavior of our observed BH mass function over the mass range probed here, although with some deviations from the overall normalization and/or slope. However, it is important to keep in mind that none of these models or simulations are able to accurately capture the detailed physics of seed formation, accretion, feedback, and dynamics, so further theoretical work is still needed, in particular when pushing to increasingly high redshifts. 

\begin{figure*}[t]
\centering
\includegraphics[angle=0,width=\textwidth]{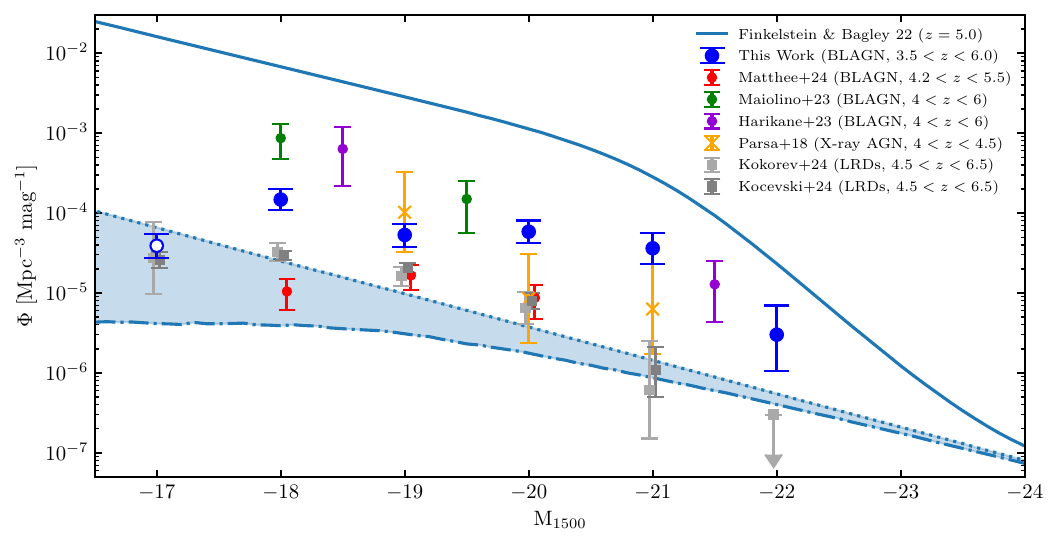}
\caption{The BLAGN+host UV luminosity function at $3.5<z<6$. We plot our UV luminosity function as blue points, and mark our $M_{1500}\geq-17$ points with hollow blue points to indicate their likely large systematic uncertainties. We show the \textit{JWST}-detected BLAGN UV luminosity functions from \cite{matthee24} (red points), \cite{Maiolino_2023} (green points), and \cite{harikane23} (purple points), the photometrically selected LRD UV luminosity functions from \cite{kokorev24} and \cite{kocevski24} as light and dark grey squares, respectively, and the x-ray detected AGN LF luminosity function from \cite{parsa18} as orange x's. We plot the best fit galaxy+AGN UV luminosity function at $z=5$ from \cite{finkelsteinbagley22} as a solid blue curve, and we shade the region between their AGN best-fit curve where the AGN UV luminosity function was permitted to turn over (blue dashed curve) and their AGN best-fit curve where the AGN UV luminosity function was fit as a power-law to faint magnitudes (blue dotted curve).}
\label{fig:uvlf}
\end{figure*}

\begin{deluxetable*}{ccccccc}[htb]
\tablecaption{BLAGN UV Luminosity Function \label{tab:uvlf}}
\tablehead{$M_{1500}$ & $\Phi$ & N & BLAGN Luminosity Fraction &\multicolumn{3}{c}{Completeness} \\  & [$10^{-6}$ Mpc$^{-3}$mag$^{-1}$] & & & Total & Obs. & Det.}
\startdata
-22.00 & $3.01^{+3.97}_{-1.95}$ & 2 & $0.129^{+0.171}_{-0.084}$ & 0.42 & 0.66 & 0.64 \\
-21.00 & $36.2^{+19.5}_{-13.4}$ & 7 & $0.128^{+0.069}_{-0.047}$ & 0.12 & 0.45 & 0.27 \\
-20.00 & $58.5^{+22.2}_{-16.6}$ & 12 & $0.052^{+0.020}_{-0.015}$ & 0.13 & 0.37 & 0.38 \\
-19.00 & $53.3^{+20.2}_{-15.2}$ & 12 & $0.019^{+0.007}_{-0.005}$ & 0.14 & 0.36 & 0.46 \\
-18.00 & $148^{+50.9}_{-38.9}$ & 14 & $0.022^{+0.008}_{-0.006}$ & 0.06 & 0.23 & 0.37 \\
-17.00 & $39^{+15.7}_{-11.6}$ & 11 & $0.002^{+0.001}_{-0.001}$ & 0.18 & 0.43 & 0.42
\enddata
\tablecomments{Columns: $M_{1500}$ is the central rest-frame UV magnitude of the UV luminosity function bin, $\Phi$ is the number density (per magnitude) of BLAGN host galaxies in each magnitude bin, N is the number of BLAGN host galaxies in each bin, Total Completeness is the average of the products of observational and line detection completeness for each object in the bin, Obs.\ Completeness is the fraction of objects in a bin that were observed by the NIRSpec MSA observation (see \S\ref{sec:completeness}), and Det.\ Completeness is the fraction of objects whose \Ha broad-lines are detected in our completeness simulations (see \S\ref{sec:completeness}).}
\end{deluxetable*}

\section{The BLAGN UV Luminosity Function}\label{sec:UVLF}
Lastly, we use our BLAGN sample to compute a UV luminosity function. To best measure UV magnitudes (at rest frame 1500~\AA{}) we revisit our \texttt{EAZY} models used in our photometric redshift fitting (see \S\ref{sec:observations}). We use \texttt{EAZY} to re-fit our BLAGN sample with redshifts fixed to the spectroscopic redshifts to produce simple SED models for each BLAGN. For each BLAGN, we then integrate the rest-frame model flux from 1450--1550~\AA{} with a top-hat filter to measure a rest-frame UV flux from which we compute an absolute rest-UV magnitude using the spectroscopic redshift. Note that we do not model or differentiate the UV flux generated by the BLAGN or by the host galaxy, thus our UV fluxes and luminosities are the combined emission from the BLAGN and host galaxy (BLAGN+host). 

We compare these UV luminosities to those measured directly from our PRISM spectroscopy (when available), and find that the EAZY-based magnitudes are $0.35^{+0.82}_{-0.53}$ magnitudes brighter than those measured directly from the PRISM data. As we only calibrate our PRISM data using \textit{JWST} NIRCam data in the F150W, F200W, F277W, F356W, F410M, and F444W filters, the PRISM data are likely poorly calibrated at rest-frame 1500~\AA{} ($0.68-1.05$~$\mu$m observed frame). Nonetheless, this offset is quite minor, and verifies the fidelity of our fitted EAZY-based magnitudes for use in the UV luminosity function.

To compensate for the effects of incompleteness, we apply the same incompleteness weightings from the BH mass function to each individual BLAGN before computing the UV luminosity function using the the same field areas, redshift range, and co-moving volume as the BH mass function. We again assume Poissonian error bars \citep{gehrels86}, with the understanding that these errors do not include systematic effects from the completeness correction. We show our resulting BLAGN+host UV luminosity function in Figure~\ref{fig:uvlf} and list the data in Table~\ref{tab:uvlf}.

We compare our UV luminosity function to the \cite{harikane23}, \cite{Maiolino_2023}, and \cite{matthee24} \textit{JWST}-detected BLAGN UV luminosity functions and find general agreement at the bright end. At moderate luminosities $(-20<M_{1500}<-18)$, our UV luminosity function lies between the higher UV luminosity functions from \cite{harikane23,Maiolino_2023} and the lower UV luminosity function from \cite{matthee24}. We also plot and compare to the photometrically identified populations of LRDs from \cite{kocevski24} and \cite{kokorev24} (grey squares); noting that while some of these are likely AGN \citep[e.g.][]{greene24,kocevski24}, the origin of the continuum emission is still very uncertain. We also compare to the best-fit galaxy+AGN UV luminosity function at $z=5$ from \citep[][solid blue curve]{finkelsteinbagley22}, their AGN best-fit curve where the AGN UV luminosity function was permitted to turn over (blue dashed curve) and their AGN best-fit curve where the AGN UV luminosity function was fit as a power-law to faint magnitudes (blue dotted curve). At all of our points at $M_{1500}\leq-18$ (where we have reasonable confidence in the completeness correction), our BLAGN+host UV luminosity function exceeds the \cite{finkelsteinbagley22} pre-\textit{JWST} AGN UV luminosity function by $\sim0.5-1.5$~dex. 

We compute the BLAGN+host fraction (the fraction of sources that host BLAGN) by dividing our BLAGN+host UV luminosity function by the full AGN+Galaxies UV luminosity function at $z=5$ from \cite{finkelsteinbagley22} in each magnitude bin. We find that at $M_{1500}=-22$, BLAGN host galaxies produce $13^{+17}_{-8}$\% of the observed UV luminosity. This fraction falls to $5.2^{+1.9}_{-1.5}$\% at $M_{1500}=-20$ and is $2.1^{+0.8}_{-0.6}$\% at $M_{1500}=-18$. Overall, the moderate to large excess of UV luminosity from BLAGN host galaxies seen in all of the \textit{JWST} BLAGN samples over the pre-\textit{JWST} data is consistent with the growing evidence seen in the \textit{JWST} era that AGN were far more prevalent in the early universe than previously observed. 

\section{Summary}\label{sec:summary}
In this work, we search for broad \Ha emission indicative of BLAGN at $3.5<z<6$ in the CEERS and RUBIES spectroscopic surveys with the goals of identifying new and previously undiscovered BLAGN, investigating the prevalence and properties of LRDs in this sample, and constructing the BH mass function. We summarize our primary results below:
\begin{enumerate}
\item{We identify a population of 62 unique BLAGN (49 of which have not previously been identified) through a systematic search for \Ha emission in the CEERS and RUBIES spectroscopic surveys. (\S\ref{sec:sample}, Figure~\ref{fig:All_Broadlines})}
\item{21 of the BLAGN are also identified as LRDs using the criteria from \cite{kocevski24}. These objects, when examined in aggregate, show broader \Ha emission profiles and a higher fraction of broad component \Ha emission than non-LRDs. (\S\ref{sec:lrds}, Figure~\ref{fig:hastack})}
\item{Most BLAGN LRDs do not have red optical slopes due solely to emission lines contamination in their broadband magnitudes. Rather, the majority of these objects are intrinsically dust reddened in the rest-frame optical. (\S\ref{sec:slopes}, Figure~\ref{fig:optvopt})}
\item{We construct the BH mass function at $3.5<z<6$ after computing robust observational and line detection completeness corrections. This BH mass function shows strong agreement with the NIRCam WFSS based BH mass function presented in \cite{matthee24}, indicating that our observational completeness correction is correct, though we extend to significantly lower BH masses.  Our results also suggest that spectroscopic BH mass functions can be constructed without explicit knowledge of the underlying spectroscopic observation selection functions. (\S\ref{sec:subBH mass function}, Figure~\ref{fig:BHmassfunction})}
\item{Our BH mass function is consistent with a variety of theoretical predictions, indicating that the observed abundance of unobscured, active black holes in the early universe is not discrepant with physically-motivated predictions of the total population of black holes. (\S\ref{sec:models}, Figure~\ref{fig:BHmassfunction_theory})}
\item{Our BH mass function shows very little structure, and most closely resembles a power-law. This lack of structure---when compared with literature and toy black hole evolution models---indicates that any memory of the underlying BH seed distribution has been lost by redshift $z\sim5$. Higher redshift observation of BLAGN will be necessary to directly probe the seed population of the first BHs and AGN. (\S\ref{sec:models}, Figure~\ref{fig:BHmassfunction_theory})}
\item{We construct the BLAGN UV luminosity function at $3.5<z<6$ using the same completeness parameters as our BH mass function. Our UV luminosity function is comparable to existing literature \textit{JWST}-detected BLAGN UV luminosity functions, and demonstrates a much higher contribution to the AGN+galaxies UV luminosity function at $3.5<z<6$ than expected in the pre-\textit{JWST} era. Our BLAGN luminosity function indicates that BLAGN account for $13^{+17}_{-8}$\% of the observed UV luminosity at $M_{1500}=-22$, $5.2^{+1.9}_{-1.5}$\% at $M_{1500}=-20$, and $2.1^{+0.8}_{-0.6}$\% at $M_{1500}=-18$. (\S\ref{sec:UVLF}, Figure~\ref{fig:uvlf})}
\end{enumerate}

This work demonstrates the power of \textit{JWST} to detect an unprecedented sample of 62 BLAGN at $3.5<z<6.8$ and to measure the black hole mass function over a wide dynamic range of mass ($\sim10^{6.5}M_{\odot}<M_{\textrm{BH}}<10^{8.5}M_{\odot}$) for the first time. Future NIRSpec MSA programs in Cycle 3 and beyond will further expand this sample, resulting in more precise BH mass functions and UV luminosity functions. The BH mass functions---especially when extended to redshifts $z>7$ by \textit{JWST} studies of H$\beta$ detected BLAGN---may place constraints on black hole seeding models and provide a better understanding of the evolution and growth of SMBHs in the early universe. Simultaneously, the BLAGN UV luminosity function may reveal the role of BLAGN in the epoch of reionization. Moreover, this sample and future samples of BLAGN will offer new insight into the evolution of galaxies and their SMBHs over the first two Gyr of cosmic time.  

\vskip 12pt
We thank the RUBIES team (including PI’s A. de Graaff and G. Brammer) for their effort designing and executing this program, and for making the data publicly available. We thank the anonymous referee for a constructive report that helped to improve this work. AJT thanks the members of the University of Texas at Austin Cosmic Frontier Center for insightful discussions during the development of this work. AJT acknowledges support from the UT Austin College of Natural Sciences, and AJT and SLF acknowledge support from STScI through award JWST-ERS-1345. RSE acknowledges generous financial support from the Peter and Patricia Gruber Foundation. YK thanks the support of the German Space Agency (DLR) through the program LEGACY 50OR2303. RA acknowledges support from project PID2023-147386NB-I00 and the Severo Ochoa grant CEX2021-001131-S funded by MCIN/AEI/10.13039/501100011033. KI acknowledges support from the National Natural Science Foundation of China (12073003, 12003003, 11721303, 11991052, and 11950410493) and the China Manned Space Project (CMS-CSST-2021-A04 and CMS-CSST-2021-A06). The Flatiron Institute is supported by the Simons Foundation. PGP-G acknowledges support from grant PID2022-139567NB-I00 funded by Spanish Ministerio de Ciencia, Innovaci\'on y Universidades MCIU/AEI/10.13039/501100011033, FEDER \textit{Una manera de hacer Europa}.

The JWST data presented in this article were obtained from the Mikulski Archive for Space Telescopes (MAST) at the Space Telescope Science Institute. The specific observations analyzed can be accessed via \dataset[doi: 10.17909/0eta-ym85]{https://doi.org/10.17909/0eta-ym85}.

\facilities{\textit{JWST}}

\software{astropy: \cite{astropy:2013,astropy:2018,astropy22}, scipy: \cite{scipy20}, emcee: \citep{emcee}, EAZY: \citep{brammer08}}

\clearpage

\begin{appendix}

\section{Description of Black Hole Toy Model}

Here, we describe in somewhat more detail the methodology of our idealized black-hole seeding and growth toy model, with results shown in the right panel of Figure~\ref{fig:BHmassfunction_theory}. We  utilize the THESAN-DARK-2 simulation output \citep{Geraldi2022,Kannan2022,Smith2022}, reading in the merger tree files across the snapshots and constructing the corresponding halo assembly histories. We either insert a seed BH of $100 M_{\odot}$, compatible with a typical light seed, or a seed of $10^4 M_{\odot}$, approximately accounting for the growth of the light seeds in minihalos that the simulation does not resolve, into the `leaves' (lowest-mass progenitors) of the trees. When exploring the heavy seed scenario, a fraction of the most massive leave-level halos is endowed with a (DCBH-like) heavy seed of $10^5 M_{\odot}$ instead. We note that the trees reach a halo mass resolution of $\sim 10^9 M_{\odot}$, somewhat more massive than the customary atomic-cooling halo scale that is often considered as the host for a heavy seed BH. The DCBH halo occupation fraction is rather uncertain, depending on the degree of fine-tuning of environmental conditions required \citep[e.g.,][]{Inayoshi2020}. We here bracket optimistic scenarios by assuming DCBH seeding fractions between 1 and 10\%. The latter value may represent an upper limit, provided by the probability that all Pop~III star formation, and the corresponding metal enrichment, can be avoided in the minihalo progenitors \citep[][]{johnson2008}. We further limit DCBH seeding to $z>10$, again reflecting the requirement of primordial (metal-free) conditions, which become increasingly rare at later times \citep{LiuBromm2020,Venditti2023}. We subsequently evolve each seed BH following a simplified recipe for the accretion rate (see below), combining aspects of Eddington-limited growth with a prescription to capture the amount of gas available in a halo. 

We assume that BH growth is less efficient in lower-mass halos because of the negative feedback from supernovae (SN) explosions. For definiteness, we use the critical halo mass suggested in \cite{Piana2021}, $M_{\rm crit}(z)=10^{11.25}M_{\odot} \times (\Omega_m(1+z)^3 + \Omega_\Lambda)^{-0.125}$, to separate an initial phase of stunted growth from the Eddington-limited growth in more massive halos. While such halo dependence may be generic, the specific value for such a critical mass depends on details of the galaxy formation process, and we here employ the \cite{Piana2021} prescription as an illustrative example only. In general, the Eddington-limited accretion rate can be written as $f_{\rm edd} \times \dot{M}_{\rm Edd}$, where $\dot{M}_{\rm Edd}$ is the Eddington rate and $f_{\rm Edd}=M_{\rm halo}/M_{\rm crit}(z)$, if $M_{\rm halo}\leq M_{\rm crit}(z)$, or 1, if $M_{\rm halo}>M_{\rm crit}(z)$ (or 1.5 when testing super-Eddington growth). Finally, we limit accretion by the available gas mass in the host halo, closely following the recipe of \cite{Piana2021}, to which we refer the reader for details, although this effect is not important for the scenarios considered here. When halos merge, we assume that their respective BHs merge instantaneously as well. We use the above recipe to evolve the BH mass through the halo merger trees until $z=5.5$, constructing a histogram of the resulting BH masses. In our accounting, we only include halos with masses above $10^{10.5} M_\odot$ at $z=5.5$ to exclude halos too faint to be represented in current \textit{JWST} surveys \citep[e.g.,][]{Jeon2023}.

\end{appendix}

\bibliography{latest.bib}

\end{document}